%% file: CanonicalBlockCov.tex
\documentclass[12pt,english]{article}
\usepackage{lmodern}

\usepackage[T1]{fontenc}
\usepackage[latin9]{inputenc}
\usepackage{geometry}
\usepackage{color}
\usepackage{float}
\usepackage{amsmath}
\usepackage{amsthm}
\usepackage{amssymb}
\usepackage{graphicx}
\usepackage{setspace}
\usepackage[authoryear]{natbib}
\makeatletter

\providecommand{\tabularnewline}{\\}

\theoremstyle{plain}
\newtheorem{thm}{\protect\theoremname}
\theoremstyle{plain}
\newtheorem{cor}{\protect\corollaryname}
\theoremstyle{plain}
\newtheorem{prop}{\protect\propositionname}

\setcitestyle{round}
\usepackage{lscape}
\usepackage{longtable}

\usepackage{graphicx}
\usepackage{tabularx}
\newcolumntype{P}[1]{>{\centering\arraybackslash}p{#1}}
\newcolumntype{Y}{>{\centering\arraybackslash}X}

\usepackage{array,booktabs,ragged2e}
\newcolumntype{C}[1]{>{\centering\arraybackslash}p{#1}}
\newcolumntype{J}[1]{>{\justify\arraybackslash}p{#1}}
\newcolumntype{R}[1]{>{\RaggedLeft\arraybackslash}p{#1}}
\newcolumntype{Q}[1]{>{\columncolor{Gray}\RaggedLeft\arraybackslash}p{#1}}
\newcolumntype{L}[1]{>{\RaggedRight\arraybackslash}p{#1}}
\newcolumntype{G}{@{\extracolsep{0.5cm}}l@{\extracolsep{0pt}}}%
\usepackage{multirow}
\usepackage{caption}

\@ifundefined{showcaptionsetup}{}{%
 \PassOptionsToPackage{caption=false}{subfig}}
\usepackage{subfig}
\AtBeginDocument{

}

\makeatother

\usepackage{babel}
\providecommand{\corollaryname}{Corollary}
\providecommand{\propositionname}{Proposition}
\providecommand{\theoremname}{Theorem}

\begin{document}
\title{A Canonical Representation of Block Matrices with Applications to
Covariance and Correlation Matrices\thanks{The second author would like to thank the Department of Statistics
and Operations Research at University of Vienna for their hospitality
during a visit in early 2020 and late 2021.}\emph{\normalsize{}\smallskip{}
}}
\author{\textbf{Ilya Archakov}$^{a}$\textbf{ and Peter Reinhard Hansen}$^{b}$\textbf{}\thanks{Address: University of North Carolina, Department of Economics, 107
Gardner Hall Chapel Hill, NC 27599-3305}\\
{\normalsize{}$^{a}$}\emph{\normalsize{}University of Vienna}\\
{\normalsize{}$^{b}$}\emph{\normalsize{}University of North Carolina
\& Copenhagen Business School}}
\date{\emph{\normalsize{}\today}}
\maketitle
\begin{abstract}
We obtain a canonical representation for block matrices. The representation
facilitates simple computation of the determinant, the matrix inverse,
and other powers of a block matrix, as well as the matrix logarithm
and the matrix exponential. These results are particularly useful
for block covariance and block correlation matrices, where evaluation
of the Gaussian log-likelihood and estimation are greatly simplified.
We illustrate this with an empirical application using a large panel
of daily asset returns. Moreover, the representation paves new ways
to regularizing large covariance/correlation matrices, test block
structures in matrices, and estimate regressions with many variables.

\bigskip{}
\end{abstract}
\begin{singlespace}
\textit{\footnotesize{}Keywords:}{\footnotesize{} Block Matrices,
Block Covariance Matrix, Block Correlation Matrix, Equicorrelation,
Covariance Regularization, Covariance Modeling, High Dimensional Covariance
Matrices, Matrix Logarithm}{\footnotesize\par}
\end{singlespace}

\noindent \textit{\small{}J}\textit{\footnotesize{}EL Classification:}{\footnotesize{}
C10; C22; C58 }{\small{}\newpage}{\small\par}

\section{Introduction}

We derive a canonical representation for a broad class of block matrices,
which includes block covariance and block correlation matrices as
special cases. The representation is a semi-spectral decomposition
of a block matrix, which is diagonalized with the exception of a single
diagonal block, whose dimension is given by the number of blocks.
The canonical representation facilitates simple computations of several
matrix functions, such as the matrix inverse, the matrix exponential,
and the matrix logarithm. Interestingly, we show that these transformations
preserve the block structure of the original matrix. Consequently,
the decomposition greatly simplifies the evaluation of Gaussian log-likelihood
functions when the covariance matrix, or the correlation matrix, has
a block structure. The canonical representation can also be used in
regressions with many regressors, instrumental variables, and dependent
variables when a block structure is appropriate.

We contribute to the literature on block correlation models by providing
simple expressions for the inverse of any (invertible) block correlation
matrix, as well as a simple expression for its determinant. The results
apply to block correlation matrices with an arbitrary number of blocks.
For block correlation matrices with two blocks, an expression for
its inverse was obtained in \citet[lemma 2.3]{EngleKelly2012}, and
related results can be found in \citet{VianaOlkin1997}.

\begin{figure}[!h]
\centering{}%
\begin{tabular}{cc}
\includegraphics[scale=0.4]{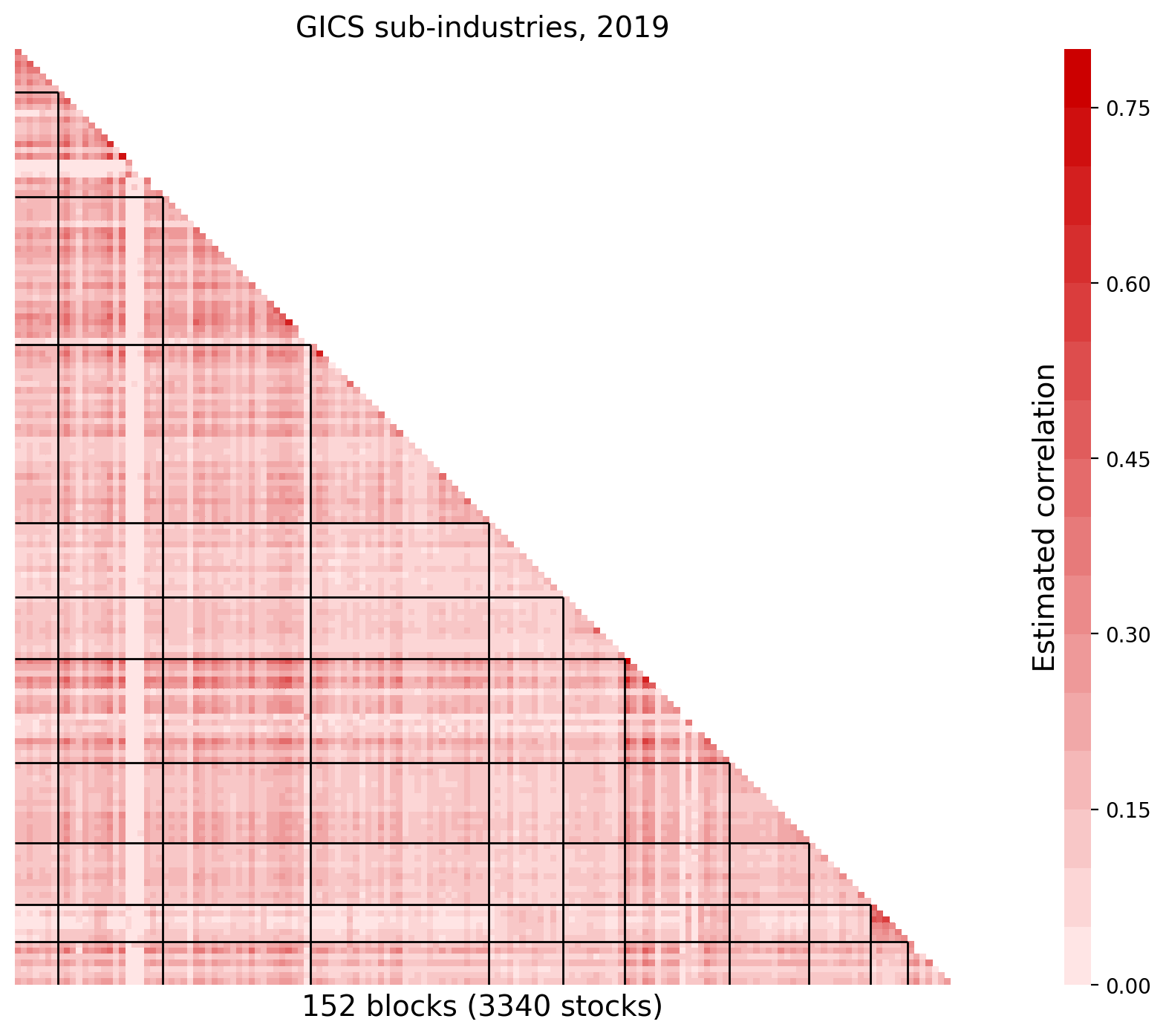} & \includegraphics[scale=0.4]{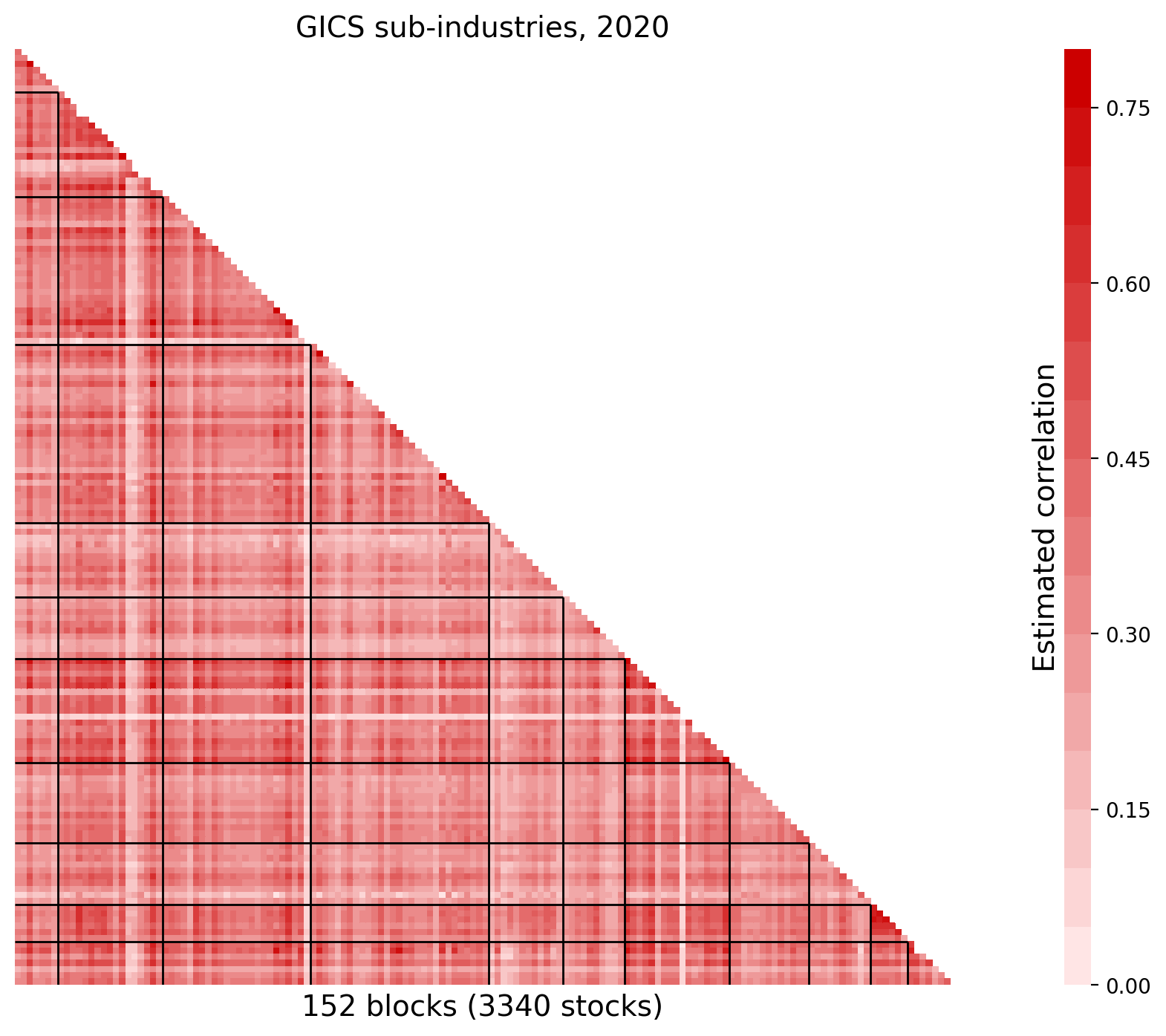}\tabularnewline
\end{tabular}\caption{Estimated $3340\times3340$ block-correlation matrices based on daily
returns for 2019 (left) and 2020 (right), where the block structure
is define by sub-industries ($K=152$ blocks).\label{fig:Preview}}
\end{figure}

We apply the block structure to estimate annual covariance matrices
of a large panel of assets using daily returns for 26 calendar years.
The block structure makes it possible to estimate and manipulate large
covariance matrices. The latter can be used to compute partial correlations.
A preview of our empirical results is presented in Figure \ref{fig:Preview},
where we use color codes to present estimated correlation matrices
for 3,340 US stocks for 2019 (left) and 2020 (right). These are estimated
with a block structure that assumes that the correlation between two
assets is defined by the sub-industries they belong to. The sub-industries
are sorted according to their Global Industry Classification Standard
(GICS) code as of 2020.\footnote{\texttt{https://en.wikipedia.org/wiki/Global\_Industry\_Classification\_Standard}}
 The solid lines indicate the boundaries of the 11 GICS sectors: Energy
(10), Materials (15), Industrials (20), Consumer Discretionary (25),
Consumer Staples (30), Health Care (35), Financials (40), Information
Technology (45), Communication Services (50), Utilities (55), and
Real Estate (60). Each of these $3340\times3340$ correlation matrices
are estimate with just 252 vectors of daily returns in 2019 and 253
daily returns in 2020. Correlations were generally larger in 2020
than in 2019. This is not surprising given the impact that the COVID-19
pandemic had on financial markets in 2020. The estimated correlations
reveal interesting differences within the sectors and sub-industries
(related to gold and other precious metals, biotechnology, and pharmaceuticals)
that are largely uncorrelated with other sub-industries. More details
are presented in Section \ref{sec:Empirical-Estimation}.

As a preview of our theoretical results, consider the $n\times n$
equicorrelation matrix,
\[
C=\left[\begin{array}{cccc}
1 & \rho & \cdots & \rho\\
\rho & 1 & \ddots & \vdots\\
\vdots & \ddots & \ddots & \rho\\
\rho & \cdots & \rho & 1
\end{array}\right],
\]
which has the eigenvalues, $1+\rho(n-1)$ and $1-\rho$, where the
latter has multiplicity $n-1$. This follows directly from the spectral
decomposition, 
\begin{equation}
Q^{\prime}CQ=D=\left[\begin{array}{cc}
1+\rho(n-1) & 0\\
0 & (1-\rho)I_{n-1}
\end{array}\right],\label{eq:QCQequi1}
\end{equation}
where $Q$ is an orthonormal matrix, i.e., $Q^{\prime}Q=I_{n}$, see
\citet{OlkinPratt:1958}. Here $I_{n}$ denotes the $n\times n$ identity
matrix. The matrix $Q$ is given by $Q=(v_{n},v_{n\bot})$, where
$v_{n}$ is the $n$-dimensional vector, $v_{n}=(\tfrac{1}{\sqrt{n}},\ldots,\tfrac{1}{\sqrt{n}})^{\prime}$,
and $v_{n\bot}$ is an $n\times(n-1)$ matrix that is orthogonal to
$v_{n}$, i.e., $v_{n\bot}^{\prime}v_{n}=0$, and orthonormal, i.e.,
$v_{n\bot}^{\prime}v_{n\bot}=I_{n-1}$.\footnote{When $n=1$, $v_{1\bot}$ is an $1\times0$ ``matrix'' and we use
the conventions: $v_{1\bot}^{\prime}v_{1\bot}=\emptyset$ (dimension
$0\times0$) and $v_{1\bot}v_{1\bot}^{\prime}=0$ (dimension $1\times1$).
This ensures that our expressions also hold in the special case, where
one or more blocks has size one.} It can now be verified that $QQ^{\prime}=I_{n}$ and $C=QQ^{\prime}CQQ^{\prime}=QDQ^{\prime}$.
In this example, $D$ is the canonical form of $C$, which is obtained
via a rotation of $C$, where the rotation matrix, $Q$, does not
depend on $\rho$.

In this paper, we derive a similar decomposition for a broad class
of block matrices, which includes block covariance matrices and block
correlation matrices as special cases. In the general case with multiple
blocks, $K\geq2$, the canonical representation does not fully disentangle
all eigenvalues. The canonical representation decomposes any block
matrix into a $K\times K$ matrix and $n-K$ real-valued eigenvalues,
where $K$ is the number of blocks. We can illustrate the general
results with a $2\times2$ block correlation matrix, $$C=
    \left[
        \begin{array}{c|c}
            C_{\rho_{11}} & \rho_{12}\mathbf{1}_{n_1\times n_2} \\ \hline 
            \bullet        & C_{\rho_{22}} 
        \end{array}
    \right]
$$where $C_{\rho_{11}}$ and $C_{\rho_{22}}$ are equicorrelation matrices
with correlations $\rho_{11}$ and $\rho_{22}$, respectively, and
dimensions $n_{1}\times n_{1}$ and $n_{2}\times n_{2}$, respectively,
and $\mathbf{1}_{n_{1}\times n_{2}}$ is the $n_{1}\times n_{2}$
matrix whose elements are all equal to one. Now define 
\[
Q=\left[\begin{array}{cccc}
v_{n_{1}} & 0 & v_{n_{1}\bot} & 0\\
0 & v_{n_{2}} & 0 & v_{n_{2}\bot}
\end{array}\right].
\]
For this correlation matrix with $2\times2$ blocks, we have the following
representation,
\begin{equation}
Q^{\prime}CQ=\left[\begin{array}{cccc}
1+\rho_{11}(n_{1}-1) & \rho_{12}\sqrt{n_{1}n_{2}} & 0 & 0\\
\rho_{12}\sqrt{n_{1}n_{2}} & 1+\rho_{22}(n_{2}-1) & 0 & 0\\
0 & 0 & (1-\rho_{11})I_{n_{1}-1} & 0\\
0 & 0 & 0 & (1-\rho_{22})I_{n_{2}-1}
\end{array}\right].\label{eq:QCQequi2}
\end{equation}
We denote the upper-left $2\times2$ matrix by $A$. In general, $A$
will be a $K\times K$ matrix, whose eigenvalues are also eigenvalues
of $C$. The general result for block matrices with $K$ blocks will
be presented in Theorem \ref{thm:Canonical}, with a structure similar
to that in (\ref{eq:QCQequi2}). Importantly, the matrix $Q$ does
not depend on the elements in block matrix, but is solely determined
by the block partition, $(n_{1},\ldots,n_{K})$, where $n=n_{1}+\cdots+n_{K}$.

The canonical representation is obtained for general block matrices,
which need not be symmetric, nor positive semidefinite. In fact, our
results are applicable to non-square matrices. Block covariance matrices
and block correlation matrices are important special cases. For block
correlation matrices, the $A$-matrix, which emerges in (\ref{eq:QCQequi2}),
was previously established in \citet{HuangYang:2010} and \citet{CadimaCalheirosPreto2010},
as we will discuss in Section \ref{sec:BlockCorrelation}. We derive
additional results for block correlation matrices that simplify various
matrix transformations and the evaluation of the Gaussian log-likelihood
function.

The rest of this paper is organized as follows. We present the main
result in Section 2, where the canonical representation is established
for a broad class of block matrices, along with related results for
some matrix functions. We also cover aspects of block structures in
regressions with many variables. In Section 3, we consider the special
case with block covariance matrices and block correlation matrices.
Many of these results are useful for maximum likelihood estimation
with a Gaussian log-likelihood function, as we show in Section 4.
In Section \ref{sec:Empirical-Estimation}, we present our empirical
analysis of block covariance matrices for a large panel of daily stock
returns. We conclude in Section 6 and all proofs are presented in
the Appendix. A separate Web Appendix with additional empirical results,
expressions for computing partial correlations from block correlation
matrices, and Matlab code used in our empirical analysis.

\section{Canonical Representation of Block Matrices}

Let $B$ be a square $n\times n$ matrix. The extension to rectangular
matrices is trivial and will be addressed towards the end of this
section. The matrix, $B$, is called a block matrix with block partition,
$n_{1},\ldots,n_{K}$, if it can be expressed as:
\[
B=\left[\begin{array}{cccc}
B_{[1,1]} & B_{[1,2]} & \cdots & B_{[1,K]}\\
B_{[2,1]} & B_{[2,2]}\\
\vdots &  & \ddots\\
B_{[K,1]} &  &  & B_{[K,K]}
\end{array}\right],
\]
where $B_{[k,l]}$ is an $n_{k}\times n_{l}$ matrix with the following
structure
\begin{equation}
B_{[k,k]}=\left[\begin{array}{cccc}
d_{k} & b_{kk} & \cdots & b_{kk}\\
b_{kk} & d_{k} & \ddots\\
\vdots & \ddots & \ddots\\
b_{kk} &  &  & d_{k}
\end{array}\right]\quad\text{and }\quad B_{[k,l]}=\left[\begin{array}{ccc}
b_{kl} & \cdots & b_{kl}\\
\vdots & \ddots\\
b_{kl} &  & b_{kl}
\end{array}\right]\quad\text{if }k\neq l,\label{eq:BlockStructure}
\end{equation}
for some constants, $d_{k}$ and $b_{kl}$, $k,l=1,\ldots,K$. So
the diagonal elements of the diagonal blocks, $B_{[k,k]}$, can take
a different value than the off-diagonal elements, whereas all elements
in an off-diagonal block, $B_{[k,l]}$, $k\neq l$, are identical.

We introduce the following notation, which relates to orthogonal projections.
Let $P_{[k,l]}=v_{n_{k}}v_{n_{l}}^{\prime}$ be the $n_{k}\times n_{l}$
matrix whose elements are all equal to $\frac{1}{\sqrt{n_{k}n_{l}}}$.
It is simple to verify that $P_{[k,m]}P_{[m,l]}=P_{[k,l]}$, and with
$k=l=m$ it follows that $P_{[k,k]}P_{[k,k]}=P_{[k,k]}$, such that
$P_{[k,k]}$ is a projection matrix. It then follows that $P_{[k,k]}^{\bot}=I_{n_{k}}-P_{[k,k]}$
is a projection matrix, and it can be verified that $P_{[k,k]}^{\bot}=v_{n_{k}\bot}v_{n_{k}\bot}^{\prime}$,
where the matrix, $v_{n\bot}$, was characterized in the introduction.

Finally, we define the $n\times n$ matrix 
\[
Q=\left[\begin{array}{cccccccc}
v_{n_{1}} & 0 & \cdots &  & v_{n_{1}\bot} & 0 & \cdots & 0\\
0 & v_{n_{2}} &  &  & 0 & v_{n_{2}\bot} &  & \vdots\\
\vdots &  & \ddots &  &  &  & \ddots\\
0 & \cdots &  & v_{n_{K}} & 0 & \cdots &  & v_{n_{K}\bot}
\end{array}\right],
\]
and observe that $Q$ is an orthonormal matrix, characterized by the
identity $Q^{\prime}Q=I$. The first $K$ columns of $Q$ can be used
to form averages within each of the $K$ blocks, whereas the remaining
columns of $Q$ capture ``differences'' within each block. The two
sets of columns span orthogonal subspaces, which correspond to distinct
components of the block decomposition. Note that $Q$ is solely defined
by the block partition, $n_{1},\ldots,n_{K}$, and it is therefore
invariant to the actual values taken by the elements in the block
matrix.
\begin{thm}
\label{thm:Canonical}Suppose that $B$ is a block matrix with block
partition $n_{1},\ldots,n_{K}$. Then
\[
B_{[k,l]}=a_{kl}P_{[k,l]}+1_{\{k=l\}}\lambda_{k}P_{[k,k]}^{\bot},\qquad\text{for}\quad k,l=1,\ldots,K,
\]
where $a_{kl}=b_{kl}\sqrt{n_{k}n_{l}}$, for $k\neq l$, $a_{kk}=d_{k}+(n_{k}-1)b_{kk}$,
and $\lambda_{k}=d_{k}-b_{kk}$. Moreover, 
\begin{equation}
B=QDQ^{\prime},\qquad\text{with}\quad D=\left[\begin{array}{cccc}
A & 0 & \cdots & 0\\
0 & \lambda_{1}I_{n_{1}-1} & \ddots & \vdots\\
\vdots & \ddots & \ddots & 0\\
0 & \cdots & 0 & \lambda_{K}I_{n_{K}-1}
\end{array}\right],\label{eq:Canonical}
\end{equation}
where $A=\{a_{kl}\}_{k,l=1}^{K}\in\mathbb{R}^{K\times K}$.
\end{thm}
The matrix $Q$ rotates $B$ into its canonical form, $D$. The first
$K$ columns of $Q$ span the eigenspace of $B$ that is associated
with the eigenvalues that $A$ and $B$ have in common. The last $n-K$
columns of $Q$ are the remaining eigenvectors of $B$.

For an arbitrary matrix, $B$, we would not expect $Q^{\prime}BQ$
to have zeroes in particular entries. The block structure imposes
a type of sparsity. Rather than imposing the sparsity on $B$, it
is imposed on $Q^{\prime}BQ$.\footnote{The repeated diagonal elements of $D$ (for $k\geq3$) implies additional
structure.}

Theorem \ref{thm:Canonical} can be used to characterize properties
of $B$ and simplifies the computation of some matrix transformations.
These include the matrix exponential, $\exp(B)=\sum_{j=0}^{-\infty}\tfrac{1}{j!}B^{j}$,
and the matrix logarithm, denoted $\log B$, which is the inverse
of $\exp(B)$. The matrix exponential and matrix logarithm are used
in spatial models, see \citet{LeSagePace:2007}, in multivariate volatility
models, see e.g., \citet{Kawakatsu:2006}, \citet{maheu-mccurdy:2011},
\citet{AsaiSo:2015}, and \citet{ArchakovHansenLundeMRG}, and play
a central role in Markov processes. The matrix logarithm is not always
well-defined, but for a positive definite symmetric matrix, $B=V\Lambda V^{\prime},$
where $\Lambda$ is the diagonal matrix with $B$'s eigenvalues, $\xi_{1},\ldots,\xi_{n}$,
we simply have $\log B=V\mathrm{diag}(\log\xi_{1},\ldots,\log\xi_{n})V^{\prime}$.
\begin{cor}
\label{cor:BlockCovMatrix}Suppose that $B$ is a block matrix as
defined above. $(i)$ The eigenvalues of $B$ are the eigenvalues
of $A$ and $\lambda_{k}=d_{k}-b_{kk}$, (the latter with multiplicity
$n_{k}-1$) $k=1,\ldots,K$, such that $\det B=\det(A)\lambda_{1}^{n_{1-1}}\cdots\lambda_{K}^{n_{K}-1}$.
$(ii)$ $B$ is invertible, if and only if $A$ is invertible and
$d_{k}\neq b_{kk}$, for all $k=1,\dots,K$. $(iii)$ The $q$-th
power of the block matrix, $B^{q}$, is well-defined whenever $A^{q}$
and $\lambda_{k}^{q}$, $k=1,\ldots,K$, are well-defined, in which
case $B^{q}$ has the same block structure as $B$, with blocks given
by
\[
B_{[k,l]}^{q}=a_{kl}^{(q)}P_{[k,l]}+1_{\{k=l\}}\lambda_{k}^{q}P_{[k,k]}^{\bot},
\]
where $a_{kl}^{q}$ is the $kl$-th element of $A^{q}$, for $k,l=1,\ldots,K$.
$(iv)$ The matrix exponential of $B$ has the same block structure
as $B$, with blocks given by
\[
\exp(B)_{[k,l]}=a_{kl}^{\exp}P_{[k,l]}+1_{\{k=l\}}e^{\lambda_{k}}P_{[k,k]}^{\bot},
\]
where $a_{kl}^{\exp}$ is the $kl$-th element of $\exp A$, for $k,l=1,\ldots,K$.
$(v)$ If $\log A$ and $\log\lambda_{k}$, $k=1,\ldots,K$, exist,
then $\log B$ has the same block structure as $B$, with blocks given
by
\[
\log(B)_{[k,l]}=a_{kl}^{\log}P_{[k,l]}+1_{\{k=l\}}\log\lambda_{k}P_{[k,k]}^{\bot},
\]
where $a_{kl}^{\log}$ is the $kl$-th element of $\log A$.
\end{cor}
It follows that $B^{q}$ is well-defined for all positive integers
of $q$, and the matrix inverse, $B^{-1}$, exists whenever $A$ is
invertible and $\lambda_{k}\neq0$, for all $k=1,\ldots,K$, in which
case $B^{q}$ is also well-defined for other negative integers of
$q$. The logarithms, $\log A$ and $\log(d_{k}-b_{kk})$, exist provided
that $A$ is invertible and $d_{k}-b_{kk}\neq0$. This may result
in a complex-valued solution to the matrix logarithm. If a real-valued
solution is required, then the conditions are that $A$ is positive
definite and that $d_{k}-b_{kk}>0$ for all $k=1,\ldots,K$.

\subsection{Block Matrices with Kronecker Representation}

Many of the expressions can be simplified further in the special case,
where all block sizes are identical, i.e., $n_{1}=n_{2}=\ldots=n_{K}=n$,
with $n=N/K$. In this situation, we have $B=A\otimes P+\Lambda\otimes P_{\bot}$,
where $P$ is the $n\times n$ matrix with $1/n$ in all entries,
$P_{\bot}=I_{n}-P$, and $\Lambda=\mathrm{diag}(\lambda_{1},\ldots,\lambda_{K})$.
In this case, it follows that $h(B)=h(A)P+h(\Lambda)P_{\bot}$, where
$h(\cdot)$ represents the matrix inverse, the matrix exponential,
or the matrix logarithm, provided these are well-defined.

\subsection{Rectangular Block Matrices}

Suppose that $B$ has blocks, $B_{[k,l]}\in\mathbb{R}^{n_{k}\times n_{l}}$,
as specified in (\ref{eq:BlockStructure}), where $k=1,\ldots,K_{1}$
and $l=1,\ldots,K_{2}$, and $K_{1}\neq K_{2}$, such that $B$ is
a non-square matrix. Set $K=\max(K_{1},K_{2})$ and suppose that $K_{1}>K_{2}$.
We conjoin zero-matrix to $B$, such that $\tilde{B}=(B,0)$ is a
square matrix with the block partition, $n_{1},\ldots,n_{K}$. Our
results apply to $\tilde{B}$, such that $\tilde{B}=QDQ^{\prime}$
has the canonical form and $B=QD\tilde{Q}^{\prime}$, where $\tilde{Q}^{\prime}$
is made up of the first $n_{1}+\cdots+n_{K_{2}}$ columns of $Q^{\prime}$.
If $K_{2}>K_{1}$, we can instead define $\tilde{B}=(B^{\prime},0)^{\prime}$,
and the results follow similarly. We will make use of rectangular
block matrices in the next subsection.

\subsection{Application to Regressions with Many Variables}

\label{subsec:Application-to-Regressions}Block matrices can be used
to impose structure in regression models with many regressors, many
instrumental variables, and many dependent variables. Consider the
standard regression model with stationary variables, $Y_{t}=\beta^{\prime}X_{t}+\varepsilon_{t}$,
$t=1,\ldots,n$ and $X_{t}\in\mathbb{R}^{q}$. If $q$ is large relative
to $n$, it may be desirable to estimate $\Sigma_{xx}=\mathbb{E}[X_{t}X_{t}^{\prime}]$
with a suitable block structure. If $Y_{t}\in\mathbb{R}^{p}$ is also
high-dimensional, one might also want to impose a block structure
on $\mathbb{E}[X_{t}Y_{t}^{\prime}]$. A similar problem arises in
regressions with many instrument, $Z_{t}$, where we can impose a
block structure on $\mathbb{E}[X_{t}Z_{t}^{\prime}]$.
\begin{thm}
\label{thm:RetangularBlockConsistent}Suppose that $(X_{t},Z_{t})$
is stationary and ergodic with finite second moment, $X_{t}\in\mathbb{R}^{q}$
and $Z_{t}\in\mathbb{R}^{m}$ with $m\geq q$. Let $\Sigma_{zx}=\mathbb{E}[Z_{t}X_{t}^{\prime}]$
and suppose that $\Sigma_{z\tilde{x}}=[\Sigma_{zx},0_{m\times(m-q)}]$
has a block structure, where $\tilde{X_{t}}=(X_{t}^{\prime},0_{1\times m-q})^{\prime}$.
Then $\Sigma_{z\tilde{x}}=QD_{z\tilde{x}}Q^{\prime}$, where the elements
of $D_{z\tilde{x}}=\mathrm{diag}(A,\lambda_{1}I_{n_{1}-1},\ldots,\lambda_{K}I_{n_{K}-1})$
can be estimated consistently, as $T\rightarrow\infty$, with 
\[
\hat{A}=\frac{1}{T}\sum_{t=1}^{T}V_{0,t}W_{0,t}^{\prime}\quad{\color{black}\text{and}}\qquad\hat{\lambda}_{k}=\frac{1}{T}\sum_{t=1}^{T}\tfrac{V_{k,t}^{\prime}W_{k,t}}{n_{k}-1},\quad k=1,\ldots,K,
\]
where $V_{t}\equiv Q^{\prime}\tilde{X}_{t}$ and $W_{t}=Q^{\prime}Z_{t}$.
The estimate of $\Sigma_{zx}$ is given by the first $q$ columns
of $\hat{\Sigma}_{z\tilde{x}}=Q\hat{D}_{z\tilde{x}}Q^{\prime}$.
\end{thm}
Theorem \ref{thm:RetangularBlockConsistent} is applicable to regression-type
estimators, such as the two-stage least squares (TSLS) estimator,
$\hat{\beta}_{\mathrm{TSLS}}=[\hat{\Sigma}_{xz}\hat{\Sigma}_{zz}^{-1}\hat{\Sigma}_{zx}]^{-1}[\hat{\Sigma}_{xz}\hat{\Sigma}_{zz}^{-1}\hat{\Sigma}_{zy}]$,
where the same, or different, block structures can be imposed on the
matrices, $\Sigma_{zx}$, $\Sigma_{zz}$, and $\Sigma_{zy}$.

Imposing a block structure entails a bias-variance trade-off, because
the block structure will induce a bias, if $\Sigma_{xz}$ does not
have the block structure. Meanwhile, a large reduction in the number
of parameters will reduce the variance of the estimator. In this context,
the unrestricted estimate, $\frac{1}{T}\sum_{t=1}^{T}Z_{t}X_{t}^{\prime}$,
is also consistent, but has a larger estimation error and many other
problems, such as those arising with many instrumental variables.
It would be interesting to develop formal tests for block structures
in order to avoid block structures that are at odds with data. The
bias-variance trade-off can motivate shrinkage methods that are based
on block structures. For instance, the use of block structures could
be combined with regularization methods, such as those in \citet{LedoitWolf:2004},
as we elaborate on in our concluding remarks.

\section{Block Correlation Matrices\label{sec:BlockCorrelation}}

A block correlation matrix is characterized by correlation coefficients
that form a block structure, where the correlation between two variables
is solely determined by the blocks to which the two variables belong.

Block correlation matrices offer a way to parameterize large covariance
matrices in a parsimonious manner. This structure is used in some
multivariate GARCH models, see \citet{EngleKelly2012} and \citet{ArchakovHansenLundeMRG}.

An $n\times n$ block correlation matrix, $C$, with $K$ blocks,
is a symmetric block matrix with, 
\begin{equation}
C_{[k,k]}=\left[\begin{array}{cccc}
1 & \rho_{kk} & \cdots & \rho_{kk}\\
\rho_{kk} & 1 & \ddots\\
\vdots & \ddots & \ddots\\
\rho_{kk} &  &  & 1
\end{array}\right]\quad\text{and for }k\neq l,\quad C_{[k,l]}=\left[\begin{array}{ccc}
\rho_{kl} & \cdots & \rho_{kl}\\
\vdots & \ddots\\
\rho_{kl} &  & \rho_{kl}
\end{array}\right],\label{eq:BlockCorrMatrix}
\end{equation}
where $\rho_{kk}$, $k=1,\ldots,K$ are within-block correlations,
and $\rho_{kl}=\rho_{lk}$, $k\neq l$, are between-block correlations,
for $k,l=1,\ldots,K$. For $C$ to be a correlation matrix, we obviously
need $\rho_{kl}\in[-1,1]$ for all $k,l=1,\ldots,K$, but this is
insufficient to guarantee a valid correlation matrix. Negative eigenvalues
can arise with some combinations of correlation coefficients, even
if these are all strictly smaller than one in absolute value.

Block equicorrelation matrices corresponds to the case where the diagonal
elements of all diagonal blocks, $B_{[k,k]}$ equal $d_{k}=1$, for
$k=1,\ldots,K$. So, Theorem \ref{thm:Canonical} fully characterizes
the set of correlation coefficients that yields a positive (semi-)
definite correlation matrix. We formulate this result as a separate
Corollary. Note that the canonical form, (\ref{eq:Canonical}), for
$C$ in (\ref{eq:BlockCorrMatrix}), yields a symmetric $A$ with
elements $a_{kl}=\rho_{kl}\sqrt{n_{k}n_{l}}$, for $k\neq l$, $a_{kk}=1+\rho_{kk}(n_{k}-1)$,
and $\lambda_{k}=1-\rho_{kk}$.
\begin{cor}[Block correlation matrices]
\label{cor:BlockPreservingTrans}Let $C$ be a block correlation
matrix. Then
\[
\det C=\det A\cdot\prod_{k=1}^{K}(1-\rho_{kk})^{n_{k}-1},
\]
such that $C$ is a non-singular block correlation matrix if and only
if $A$ is positive definite and $|\rho_{kk}|<1$. In this case, both
$C^{-1}$ and $\log C$ have the same block structure as $C$, with
blocks given by
\[
C_{[k,l]}^{-1}=a_{kl}^{\#}P_{[k,l]}+1_{\{k=l\}}\tfrac{1}{1-\rho_{kk}}P_{[k,k]}^{\bot},
\]
and
\[
\log(C)_{[k,l]}=\tilde{a}_{kl}P_{[k,l]}+1_{\{k=l\}}\log(1-\rho_{kk})P_{[k,k]}^{\bot},
\]
respectively, where $a_{kl}^{\#}$ is the $kl$-th element of $A^{-1}$
and $\tilde{a}_{kl}$ is the $kl$-th element of $\log A$.
\end{cor}
For $C$ in (\ref{eq:BlockCorrMatrix}) to be a correlation matrix
(possibly singular), we need that $A$ is positive semidefinite and
that $|\rho_{kk}|\leq1$, and Corollary \ref{cor:BlockPreservingTrans}
characterizes the set of positive definite block equicorrelation matrices.
The additional requirements are that $A$ is positive definite and
$|\rho_{kk}|<1$, $k=1,\ldots,K$.

In this context with block correlation matrices, the expression for
$A$ was previously obtained in \citet[proposition 5]{HuangYang:2010}
and in \citet[theorem 3.1]{CadimaCalheirosPreto2010}. \citet{HuangYang:2010}
focused on computational issues, which might explain that their paper
is overlooked in much of the literature.\footnote{We were, until recently, also unaware of the results in \citet{HuangYang:2010}
and \citet{CadimaCalheirosPreto2010}. An anonymous referee (on a
different paper than the present one) directed us to \citet{RoustantDeville2017}
and we subsequently discovered the more detailed results in \citet{HuangYang:2010}
and \citet{CadimaCalheirosPreto2010}. Some of their results, e.g.,
\citet[eq. 6]{HuangYang:2010}, were rediscovered in \citet{RoustantDeville2017},
who do not cite \citet{HuangYang:2010} or \citet{CadimaCalheirosPreto2010}.
In fact, none of the papers, \citet{CadimaCalheirosPreto2010}, \citet{HuangYang:2010},
\citet{EngleKelly2012}, and \citet{RoustantDeville2017} cite any
of the other papers listed here. The results in \citet{RoustantDeville2017}
appear to have been absorbed and extended in \citet{RoustantEtAl2020}.} Their results add valuable insight about the block-DECO model by
\citet{EngleKelly2012}. For instance, their results provide a simple
way to evaluate if a block correlation matrix is positive definite
(or semidefinite). The expression for the determinant of a correlation
matrix in Corollary \ref{cor:BlockPreservingTrans} is a simple implication
of the eigenvalues derived in \citet{HuangYang:2010} and \citet{CadimaCalheirosPreto2010},
whereas the expressions for the inverse and logarithmically transformed
correlation matrices are new, and so is our results on the preservation
of block structures for certain matrix functions.

\section{Applications of the Canonical Representation to Gaussian Log-Likelihood}

In this section, we focus on covariance and correlation matrices for
normally distributed random variables. We derive simplified expressions
for the corresponding log-likelihood functions, which greatly reduce
the computational burden when $n$ is large relative to $K$. We derive
the maximum likelihood estimator and provide a simple expression for
the first derivatives of the log-likelihood function with respect
to the unknown parameters (the scores).

We will follow the conventional notation for covariances and variances,
we write $\sigma_{kl}$ in place of $b_{kl}$, $k,l=1,\ldots,K$,
and $\sigma_{k}^{2}$ in place of $d_{k}$, $k=1,\ldots,K$. Similarly,
for correlation matrices we write $\rho_{kl}$ in place of $b_{kl}$,
and have $d_{k}=1$.

The density function for the multivariate Gaussian distribution with
mean zero and an $n\times n$ covariance matrix, $\Sigma$, is $f(x)=(2\pi)^{-\frac{n}{2}}(\det\Sigma)^{-\frac{1}{2}}\exp(-\tfrac{1}{2}x^{\prime}\Sigma^{-1}x)$.
Suppose that $\Sigma$ has the block structure given by $(n_{1},\ldots,n_{K})$,
and let $\Sigma=QDQ^{\prime}$ be its canonical representation. The
corresponding log-likelihood function (multiplied by $-2$) can be
expressed as
\[
-2\ell=n\log2\pi+\log\det D+X^{\prime}QD^{-1}Q^{\prime}X,
\]
where $D=\mathrm{diag}(A,\lambda_{1}I_{n_{1}-1},\ldots,\lambda_{K}I_{n_{K}-1})$,
with $\lambda_{k}=\sigma_{k}^{2}-\sigma_{k,k}$, and 
\[
a_{kl}=\begin{cases}
\sigma_{k}^{2}+(n_{k}-1)\sigma_{k,k} & \text{for }k=l,\\
\sigma_{k,l}\sqrt{n_{k}n_{l}} & \text{for }k\neq l.
\end{cases}
\]
So, if we define $Y=(y_{0}^{\prime},y_{1}^{\prime},\ldots,y_{K}^{\prime})^{\prime}=Q^{\prime}X$,
where $y_{0}$ is $K$-dimensional and $y_{k}$ is $n_{k}-1$ dimensional,
$k=1,\ldots,K$, then it follows that
\begin{equation}
-2\ell=n\log2\pi+\log\det A+y_{0}^{\prime}A^{-1}y_{0}+\sum_{k=1}^{K}\left((n_{k}-1)\log\lambda_{k}+\tfrac{y_{k}^{\prime}y_{k}}{\lambda_{k}}\right).\label{eq:-2logL}
\end{equation}
This expression shows that the block structure yields a considerable
simplification for log-likelihood evaluation. Instead of inverting
the $n\times n$ matrix $\Sigma$ and computing $\det\Sigma$, it
suffices to invert the smaller $K\times K$ matrix, $A$, and evaluate
$\det A$. Moreover, the maximum likelihood estimator based on a random
sample, $X_{1},\ldots,X_{N}$, is easily expressed in terms of the
transformed variables, $Y_{1}=Q^{\prime}X_{1},\ldots,Y_{N}=Q^{\prime}X_{N}$,
as formulated in the following theorem.
\begin{thm}
\label{thm:BlockVarianceMLE}Suppose that $X_{t}\sim iidN(0,\Sigma)$,
$t=1,\ldots,T$ where $\Sigma$ is a block covariance matrix with
block partition, $n_{1},\ldots,n_{K}$. Define the transformed variables,
$Y_{t}=Q^{\prime}X_{t}=(y_{0,t}^{\prime},y_{1,t}^{\prime},\ldots,y_{K,t}^{\prime})^{\prime}$,
where $y_{0,t}\in\mathbb{R}^{K}$ and $y_{k,t}\in\mathbb{R}^{n_{k}-1}$,
$k=1,\ldots,K$.

Then $\text{\ensuremath{\hat{\Sigma}=Q\hat{D}Q^{\prime}}}$ is the
maximum likelihood estimator of $\Sigma$, where $\hat{D}=\mathrm{diag}(\hat{A},\hat{\lambda}_{1}I_{n_{1}-1},\allowbreak\ldots,\allowbreak\hat{\lambda}_{K}I_{n_{K}-1})$
with
\[
\hat{A}=\frac{1}{T}\sum_{t=1}^{T}y_{0,t}y_{0,t}^{\prime}\quad\text{and}\qquad\hat{\lambda}_{k}=\frac{1}{T}\sum_{t=1}^{T}\tfrac{y_{k,t}^{\prime}y_{k,t}}{n_{k}-1},\quad k=1,\ldots,K.
\]
\end{thm}
The maximum likelihood estimates of the individual parameters can
be obtained directly from $\hat{A}$ and $\hat{\lambda}_{k}$, $k=1,\ldots,K$.
From the definition of $A$ it follows that $\hat{\sigma}_{k,l}=\hat{a}_{kl}/\sqrt{n_{k}n_{l}}$
for $k\neq l$, and for $k=l$, we have $\hat{\sigma}_{k,k}=(\hat{a}_{kk}-\hat{\lambda}_{k})/n_{k}$
and $\hat{\sigma}_{k}^{2}=\hat{\lambda}_{k}+\hat{\sigma}_{k,k}=\tfrac{1}{n_{k}}\hat{a}_{kk}+\tfrac{n_{k}-1}{n_{k}}\hat{\lambda}_{k}$.\footnote{This follows from the definitions, $\sigma_{k}^{2}=\lambda_{k}+\sigma_{kk}$
and $a_{kk}=\sigma_{k}^{2}+(n_{k}-1)\sigma_{kk}$, such that $a_{kk}=\lambda_{k}+\sigma_{kk}+(n_{k}-1)\sigma_{kk}=\lambda_{k}+n_{k}\sigma_{kk}$,
and the invariance of the maximum likelihood estimator.}

In the special case where a block has size one, we have $\Sigma_{[k,k]}=\sigma_{k}^{2}$
and $\sigma_{k,k}$ is undefined. In this situation, the corresponding,
$y_{k,t}$, is also undefined, and hence, so is $\hat{\lambda}_{k}$.
Yet the expressions for the maximum likelihood estimators continue
to be valid, including the expression for $\hat{\Sigma}$ in Theorem
\ref{thm:BlockVarianceMLE}. If $n_{k}=1$, then $\hat{\sigma}_{k}^{2}=\hat{a}_{kk}$,
while the expression for $\hat{\sigma}_{k,k}$ is redundant and can
be ignored.

Estimation when the correlation matrix is assumed to have a block
structure, as opposed to the covariance matrix having a block structure,
is more convoluted. A block correlation matrix is entirely given by
the $A$-matrix, because the eigenvalues $\lambda_{1},\ldots,\lambda_{K}$
are given from $A$. Below we use the notation, $\mathcal{I}_{k}\equiv\{i_{k-1}+1,\ldots,i_{k}\}$,
with $i_{k}=\sum_{j=1}^{k}n_{j}$, which contains the $n_{k}$ indices
associated with the $k$-th block.
\begin{cor}
\label{cor:BlockCorrelationMLE}Suppose that $X_{t}\sim iidN_{n}(0,\Sigma)$,
$t=1,\ldots,T$, where $\Sigma=\Lambda_{\sigma}C\Lambda_{\sigma}$
with $\Lambda_{\sigma}=\mathrm{diag}(\sigma_{1},\ldots,\sigma_{n})$
and $C$ is a block correlation matrix with block partition, $n_{1},\ldots,n_{K}$.
The maximum likelihood estimates of $\sigma_{1}^{2},\ldots,\sigma_{n}^{2}$
satisfy
\begin{equation}
\frac{1}{n_{k}}\sum_{i\in\mathcal{I}_{k}}\frac{s_{i}^{2}}{\tilde{\sigma}_{i}^{2}}=1,\qquad k=1,\ldots,K,\label{eq:MLEsigmaBlockC}
\end{equation}
where $s_{i}^{2}=T^{-1}\sum_{t=1}^{T}X_{i,t}^{2}$, for $i=1,\ldots,n$.
Let $\tilde{X}_{i,t}=X_{i,t}/\tilde{\sigma}_{i}$ and define $\tilde{Y}_{t}=Q^{\prime}\tilde{X}_{t}=(\tilde{y}_{0,t}^{\prime},\tilde{y}_{1,t}^{\prime},\ldots,\tilde{y}_{K,t}^{\prime})^{\prime}$
where $\tilde{y}_{0,t}\in\mathbb{R}^{K}$ and $\tilde{y}_{k,t}\in\mathbb{R}^{n_{k}-1}$,
$k=1,\ldots,K$.

The maximum likelihood estimator of $C$ is $\tilde{C}=Q\tilde{D}Q^{\prime}$,
where $\tilde{D}=\mathrm{diag}(\tilde{A},\tilde{\lambda}_{1}I_{n_{1}-1},\allowbreak\ldots,\allowbreak\tilde{\lambda}_{K}I_{n_{K}-1})$
with
\[
\tilde{A}=\frac{1}{T}\sum_{t=1}^{T}\tilde{y}_{0,t}\tilde{y}_{0,t}^{\prime}\quad\text{and}\qquad\tilde{\lambda}_{k}=\frac{n_{k}-\tilde{a}_{kk}}{n_{k}-1},\quad k=1,\ldots,K,
\]
where $\tilde{a}_{kk}$ is the $k$-th diagonal element of $\tilde{A}$.
\end{cor}
Thus, the estimate of $D$ can be obtained solely from the $K\times K$
matrix $\tilde{A}$. For the individual correlations we have $\tilde{\rho}_{k,l}=\tilde{a}_{kl}/\sqrt{n_{k}n_{l}}$,
for $k\neq l$, and $\tilde{\rho}_{k,k}=(\tilde{a}_{kk}-1)/(n_{k}-1)$,
for $k=l$. Corollary \ref{cor:BlockCorrelationMLE} does not fully
specify the maximum likelihood estimates of $(\sigma_{1}^{2},\ldots,\sigma_{n}^{2})$,
but these are given in the proof in implicit form, and some additional
details are stated immediately after the proof.

A simple and consistent estimator, as $T\rightarrow\infty$, is to
set $\hat{\sigma}_{i}^{2}=s_{i}^{2}$, $i=1,\ldots,n$, which satisfy
(\ref{eq:MLEsigmaBlockC}), and compute the corresponding $\hat{C}$,
with the expression for $A$ and $\lambda_{k}$, $k=1,\ldots,K$,
which maximizes the log-likelihood function subject to $\sigma_{i}^{2}=s_{i}^{2}$,
$i=1,\ldots,n$. We refer to this estimator as the two-stage estimator.

The score of the log-likelihood function is often of separate interest.
For instance, the score is used for the computation of robust standard
errors, in Lagrange multiplier tests, in tests for structural breaks,
see e.g., \citet{Nyblom89}, and in dynamic models with time-varying
parameters (the so-called score-driven models), see \citet{CrealKoopmanLucasGAS_JAE}.
So we provide the expressions for the score in this context with a
block covariance matrix.

Suppose that $\Sigma$ is a block covariance matrix and let $\Sigma=QDQ^{\prime}$
be its canonical representation. Because $Q$ is entirely given by
the block partition $(n_{1},\ldots,n_{K})$, and does not depend on
the unknown parameters in $\Sigma$, the expressions for the partial
derivatives are relatively simple.
\begin{prop}
\label{prop:Score}Let $\Sigma=QDQ$ be the canonical representation
of $\Sigma$. Then $\partial(-2\ell)/\partial A=M=A^{-1}-A^{-1}y_{0}y_{0}^{\prime}A^{-1}$
and, for $k=1,\ldots,K$, we have
\begin{eqnarray*}
\frac{\partial(-2\ell)}{\partial\sigma_{k}^{2}} & = & M_{k,k}+\left(\tfrac{n_{k}-1}{\lambda_{k}}-\tfrac{y_{k}^{\prime}y_{k}}{\lambda_{k}^{2}}\right)\\
\frac{\partial(-2\ell)}{\partial\sigma_{kk}} & = & (n_{k}-1)M_{k,k}-\left(\tfrac{n_{k}-1}{\lambda_{k}}-\tfrac{y_{k}^{\prime}y_{k}}{\lambda_{k}^{2}}\right),
\end{eqnarray*}
and, for $i\neq j$, we have $\frac{\partial(-2\ell)}{\partial\sigma_{ij}}=2\sqrt{n_{k}n_{l}}M_{i,j}.$
\end{prop}
The hessian could be derived similarly. In some applications, it might
be preferable to parametrize the block covariance matrix with $A$
and ($\lambda_{1},\ldots,\lambda_{K})$. In this case, one can use
$\partial(-2\ell)/\partial A=M$, and $\partial(-2\ell)/\partial\lambda_{k}=\tfrac{n_{k}-1}{\lambda_{k}}-\tfrac{y_{k}^{\prime}y_{k}}{\lambda_{k}^{2}}$,
for $k=1,\ldots,K$.

\section{Empirical Estimation of Block Correlation Matrices}

\label{sec:Empirical-Estimation}We proceed to illustrate how high-dimensional
covariance matrices with a block structure are straightforward to
estimate in practice. We estimate block correlation matrices for a
large panel of daily asset returns for each of the years from 1995
to 2020. We include all stocks from the Center for Research in Security
Prices (CRSP) database that could be matched with a unique permanent
number (PERMNO from the Compustat data). Stocks with missing observations
in a calendar year were excluded from the estimation in that calendar
year. Across years, we have between $n=3,340$ and $n=6,637$ stocks,
with an average of 4,446 stocks per year. Each calendar year has $T\simeq250$
daily returns, which we use to estimate the $n\times n$ correlation
matrix for that year.

The objective of this empirical application is to demonstrate that
high-dimensional covariance matrices can be estimated with relatively
few observations once a block structure is imposed, and that the canonical
representation makes it simple to obtain consistent estimates and
evaluate the Gaussian log-likelihood function. Because variances and
covariances vary over time, our estimates reflect correlations implied
by the average covariance matrices over each calendar year, rather
than an accurate description of the data generating process.

In our analysis, we inspect five nested block structures for the correlation
matrix, where the equicorrelation structure ($K=1$) is the simplest
and most restrictive model. The other four correlation models use
block structures defined by GICS Sectors, Groups, Industries, and
Sub-Industries. The numbers of blocks are increasing from Sectors
to Sub-Industries, but may change from year to year within a given
category. These correspond to $K=11$ for Sectors, and across years
$K$ ranges between 24 and 26 for Groups, between 69 and 76 for Industries,
and between 152 and 182 for Sub-Industries.

We estimate the canonical correlation matrix for each calendar year
using the two-stage estimator. Hence, the individual variances are
estimated with the sample variances, $\hat{\sigma}_{i}^{2}=T^{-1}\sum_{t=1}^{T}r_{i,t}^{2}$,
where the daily returns, $r_{i,t}$, are adjusted for dividends and
stock splits. Then the block correlation matrix is estimated from
standardized returns, $\hat{z}_{i,t}=r_{i,t}/\hat{\sigma}_{i}$, using
Corollary \ref{cor:BlockCorrelationMLE}, and we obtain $\hat{\rho}_{k,l}=\hat{a}_{kl}/\sqrt{n_{k}n_{l}}$,
for $k\neq l$, and $\hat{\rho}_{k,k}=(\hat{a}_{kk}-1)/n_{k}$, where
$\hat{A}=\frac{1}{T}\sum_{t=1}^{T}\hat{y}_{0,t}\hat{y}_{0,t}^{\prime}$
and the $k$-th element of $\hat{y}_{0,t}$ is given by $\frac{1}{\sqrt{n_{k}}}\sum_{i\in\mathcal{I}_{k}}\hat{z}_{i,t}$.
So, the entire $n\times n$ covariance matrix with the block correlation
structure is estimated by $n$ (univariate) variances and the $K\times K$
matrix, $\hat{A}$. The unrestricted estimate would obviously be singular
because the dimension, $n$, is an order of magnitude larger than
$T$ in all calendar years. The block assumption imposes enough structure
for $\hat{C}$ to be invertible, which requires an inverse of the
$K\times K$ matrix, $\hat{A}$.
\begin{table}[!h]
\begin{centering}
\bigskip{}
\input{crisp_mle_stat_4.tex}
\par\end{centering}
\caption{Summary statistics for the estimated block correlation matrices.\label{tab:Estimates}}
\end{table}

To conserve space, we only provide the most detailed estimation results
for the last two calendar years, 2019 and 2020, and partial correlations
will be presented for six calendar years. Detailed results for all
26 calendar years (1995-2018) are presented in the Web Appendix, \citet{ArchakovHansen:BlockAppendix}.
Comparing 2019 and 2020 is interesting because some effects of the
COVID-19 pandemic can be observed in 2020. Both years have the same
number of assets, $n=3,340$ assets, and the same number of blocks
for all five block structures.

In Table \ref{tab:Estimates}, we report the range of estimated correlations
for each of the five types of block structure and for both calendar
years, 2019 and 2020. The range of estimated correlations was larger
in 2020 than in 2019, and it increases as the number of blocks increases.
The latter is expected because the number of distinct correlation
coefficients in $C$ increases as $K$ increases. Each estimated correlation
represents an average correlation, subject to both time averaging
(over a calendar year) and cross-sectional averaging within the corresponding
sector/group/industry/sub-industry. We also report the log-likelihood
function (scaled by $-2/(nT)$) evaluated at the parameter estimates,
and the corresponding value of the Bayesian Information Criterion
(BIC).\footnote{These are approximate BIC statistics, because they are based on the
two-estimator.} The minimum BIC is obtained with a block structures based on Groups
in both 2019 and 2020.\footnote{The BIC adds the penalty $p\log(nT)$ to $-2\ell$, where $p$ is
the number of free parameters. For comparison, the AIC, which uses
the penalty $2p$, selects the most general specification in both
years.} The last column reports the number $K(K+1)/2$ of unique correlations
within a block structure with $K$ blocks, and while this number increases
rapidly with $K$, the gains in the log-likelihood are relatively
modest. Consequently, the BIC increases substantially when blocks
are defined by sub-industries.
\begin{figure}[!ph]
\begin{centering}
\subfloat[Sector correlation structure ($K=11$ blocks)]{\centering{}%
\begin{tabular}{cc}
\includegraphics[scale=0.4]{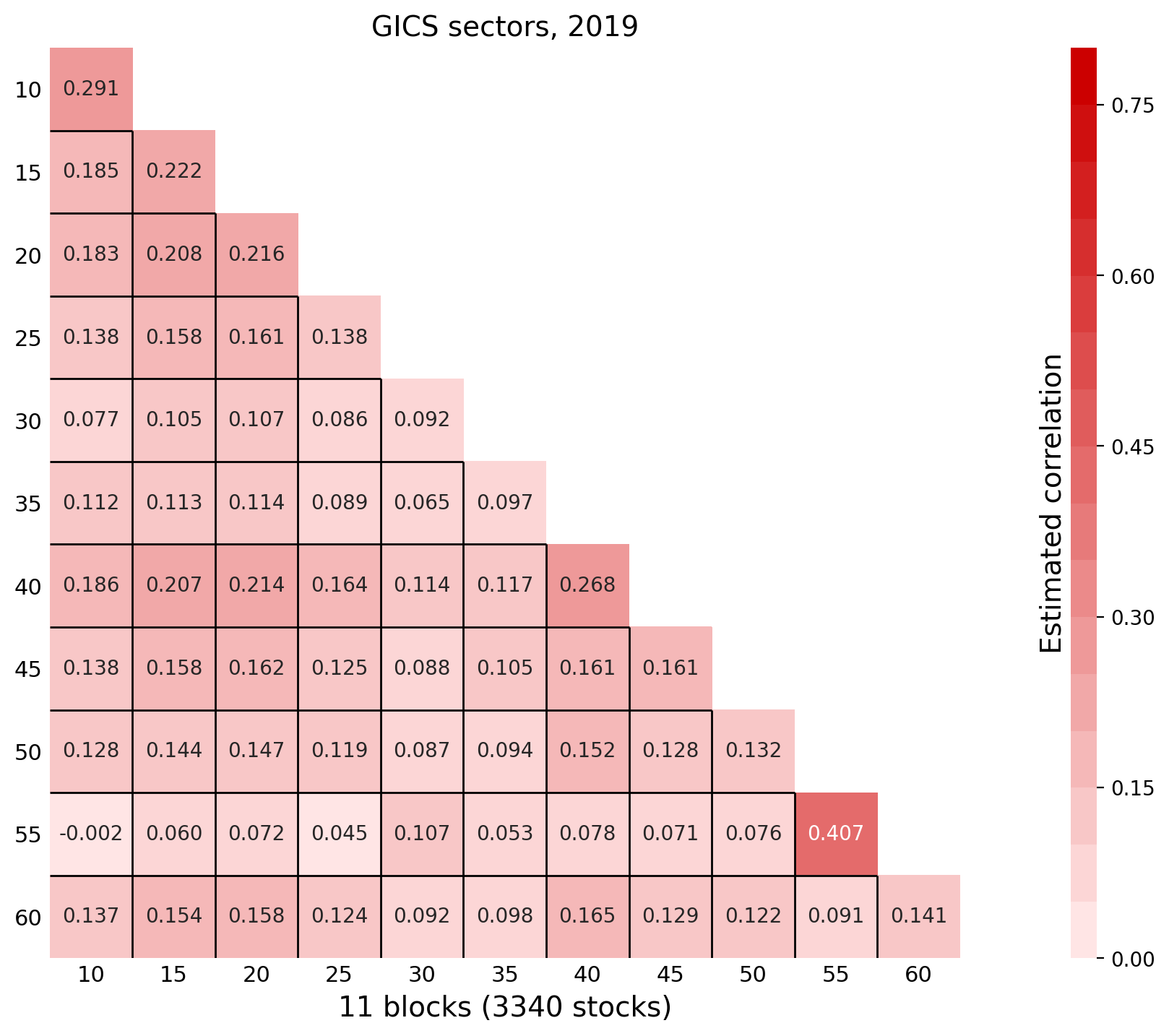} & \includegraphics[scale=0.4]{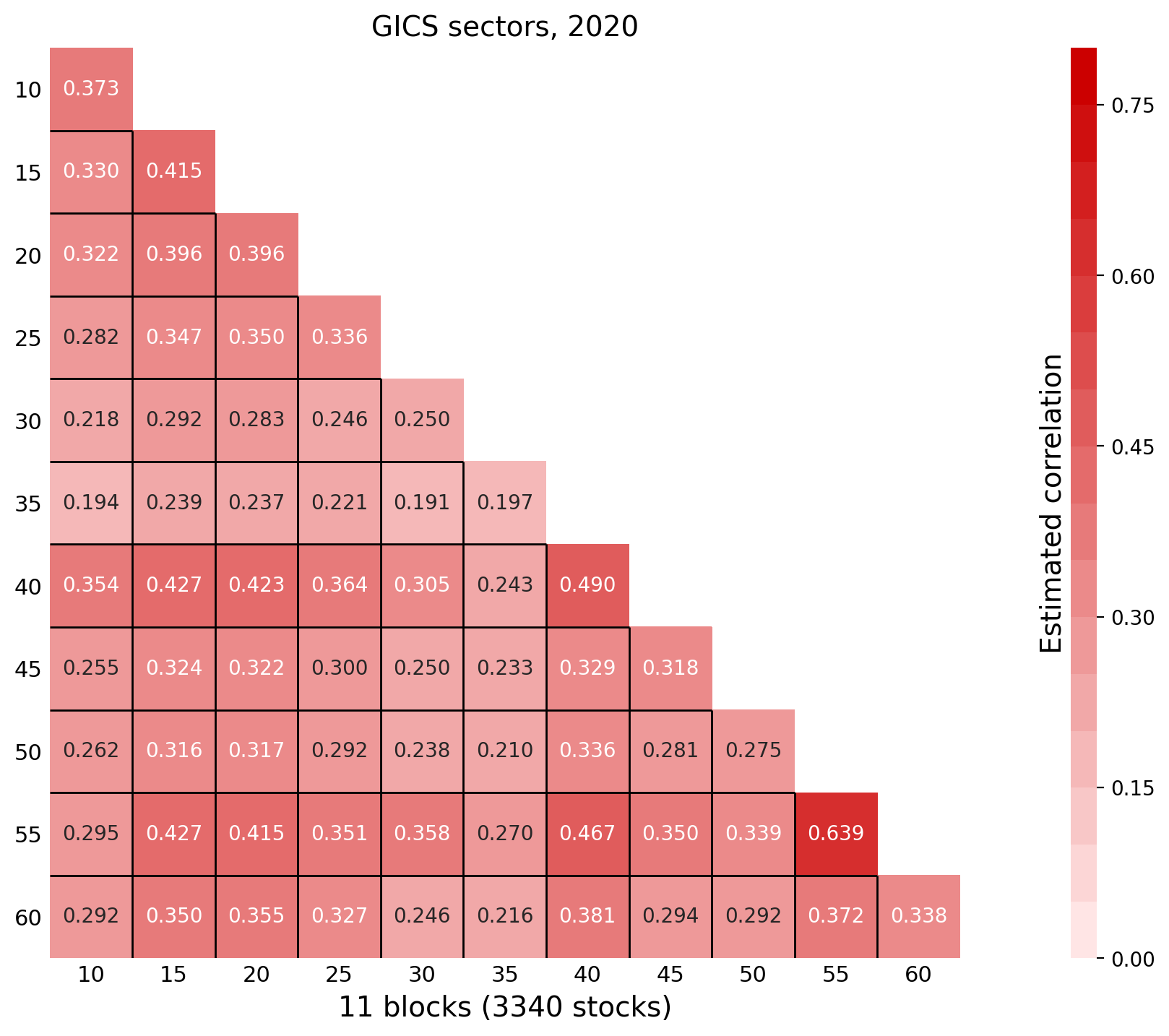}\tabularnewline
\end{tabular}}
\par\end{centering}
\begin{centering}
\subfloat[Group correlation structure ($K=24$ blocks)]{\centering{}%
\begin{tabular}{cc}
\includegraphics[scale=0.4]{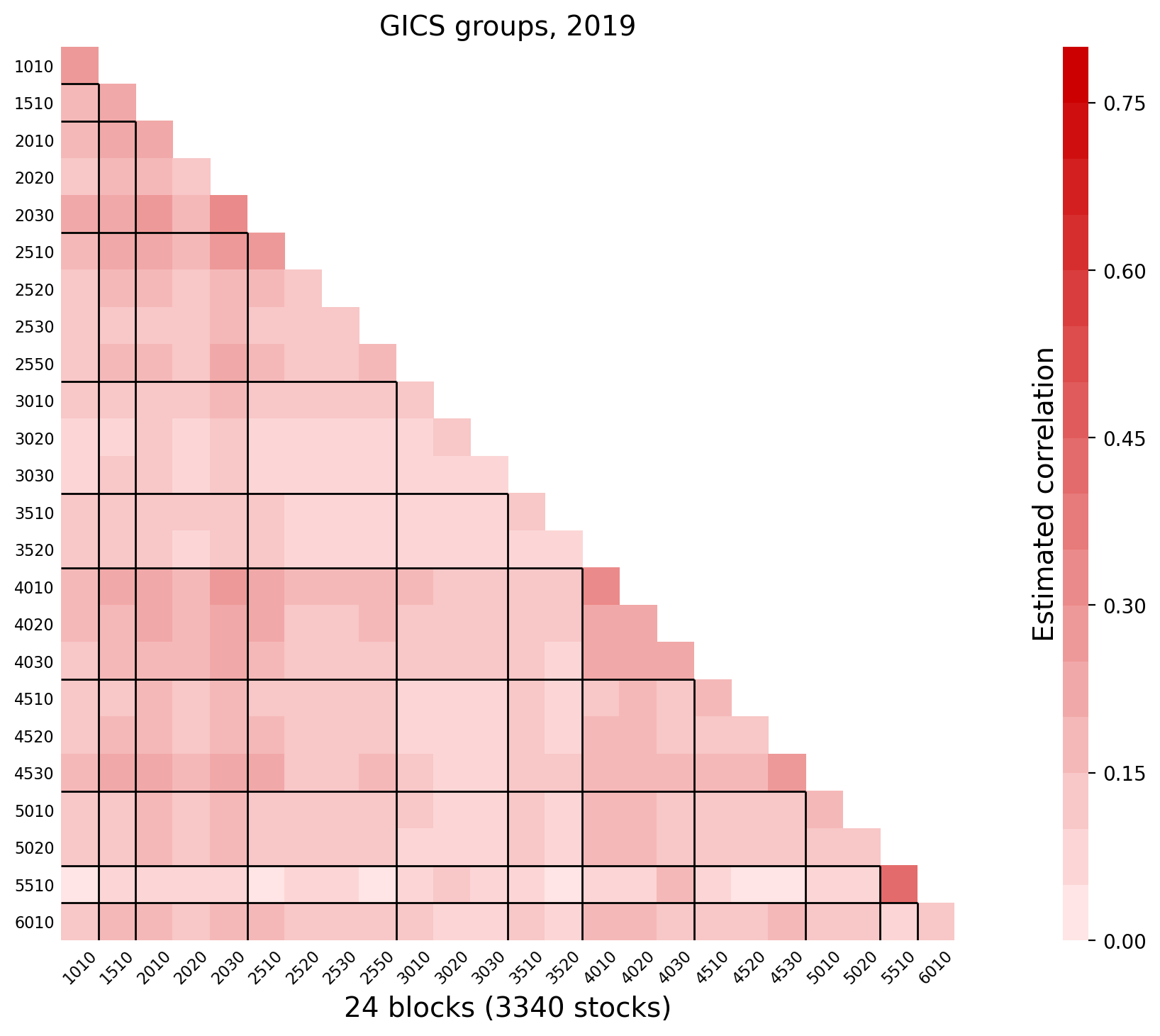} & \includegraphics[scale=0.4]{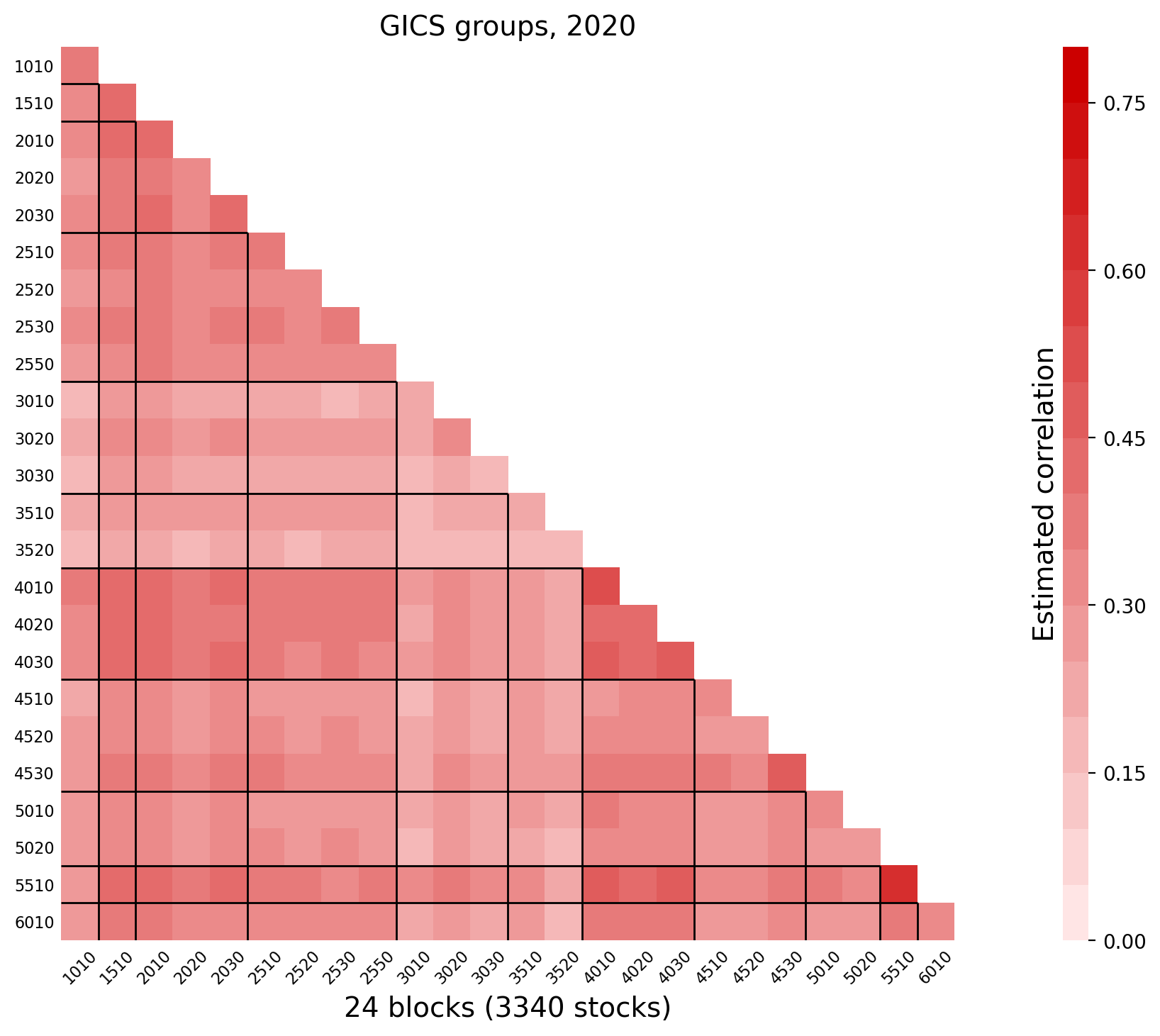}\tabularnewline
\end{tabular}}
\par\end{centering}
\caption{Estimated correlations for block structure based on Sectors (a), Groups
(b), Industries (c) and Sub-industries (d). Left panels are the estimates
for 2019 based on 252 daily returns and the right panels are for 2020
based on 253 daily returns from 2020. The block of asset are listed
according to their GICS classification code, and the black solid lines
indicated the boundaries of the 11 sectors.\label{fig:BlockCorrelationsMatrices}}
\end{figure}
\begin{figure}[!ph]
\begin{centering}
\ContinuedFloat\subfloat[Industry correlation structure ($K=69$ blocks)]{\centering{}%
\begin{tabular}{cc}
\includegraphics[scale=0.4]{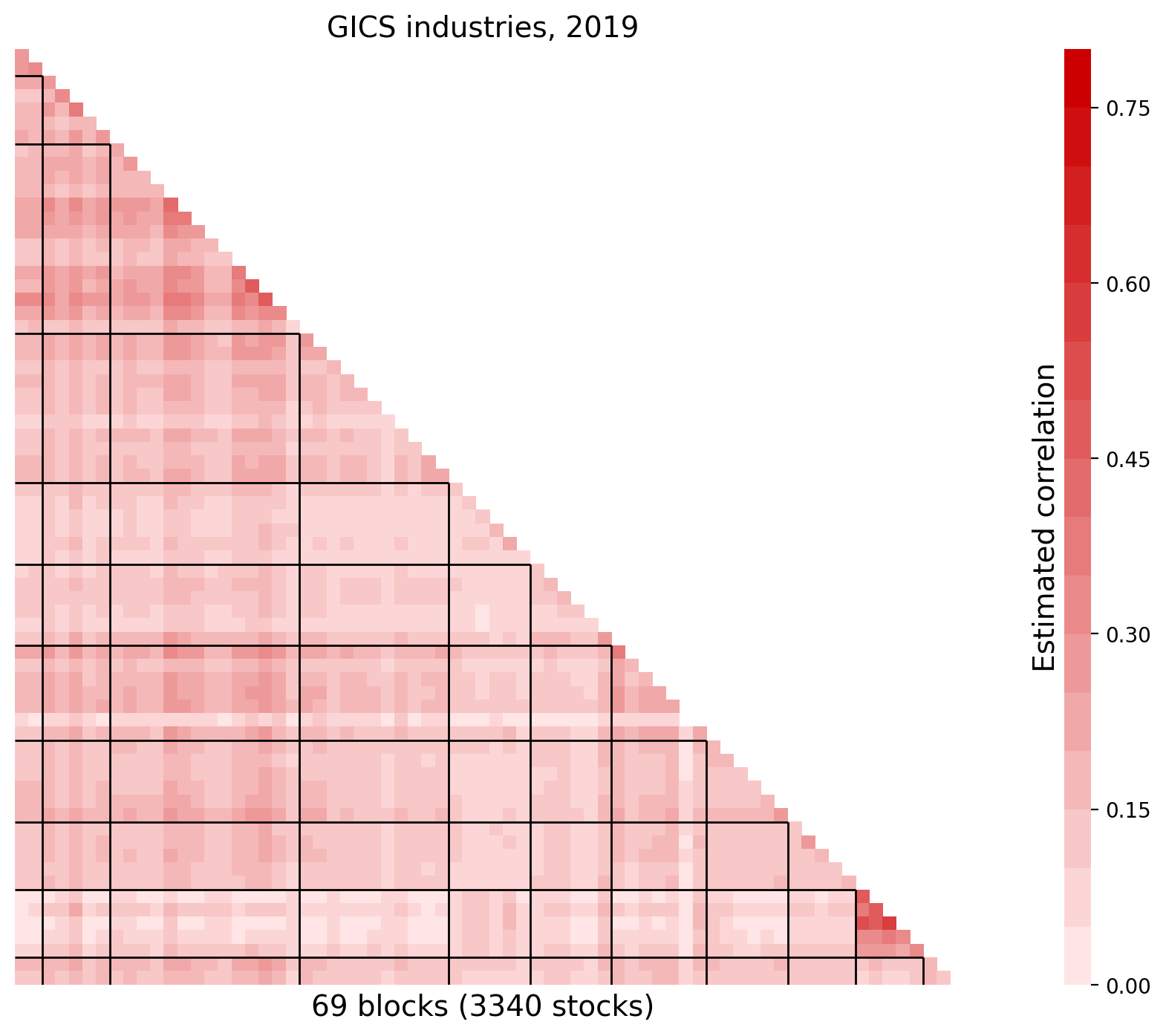} & \includegraphics[scale=0.4]{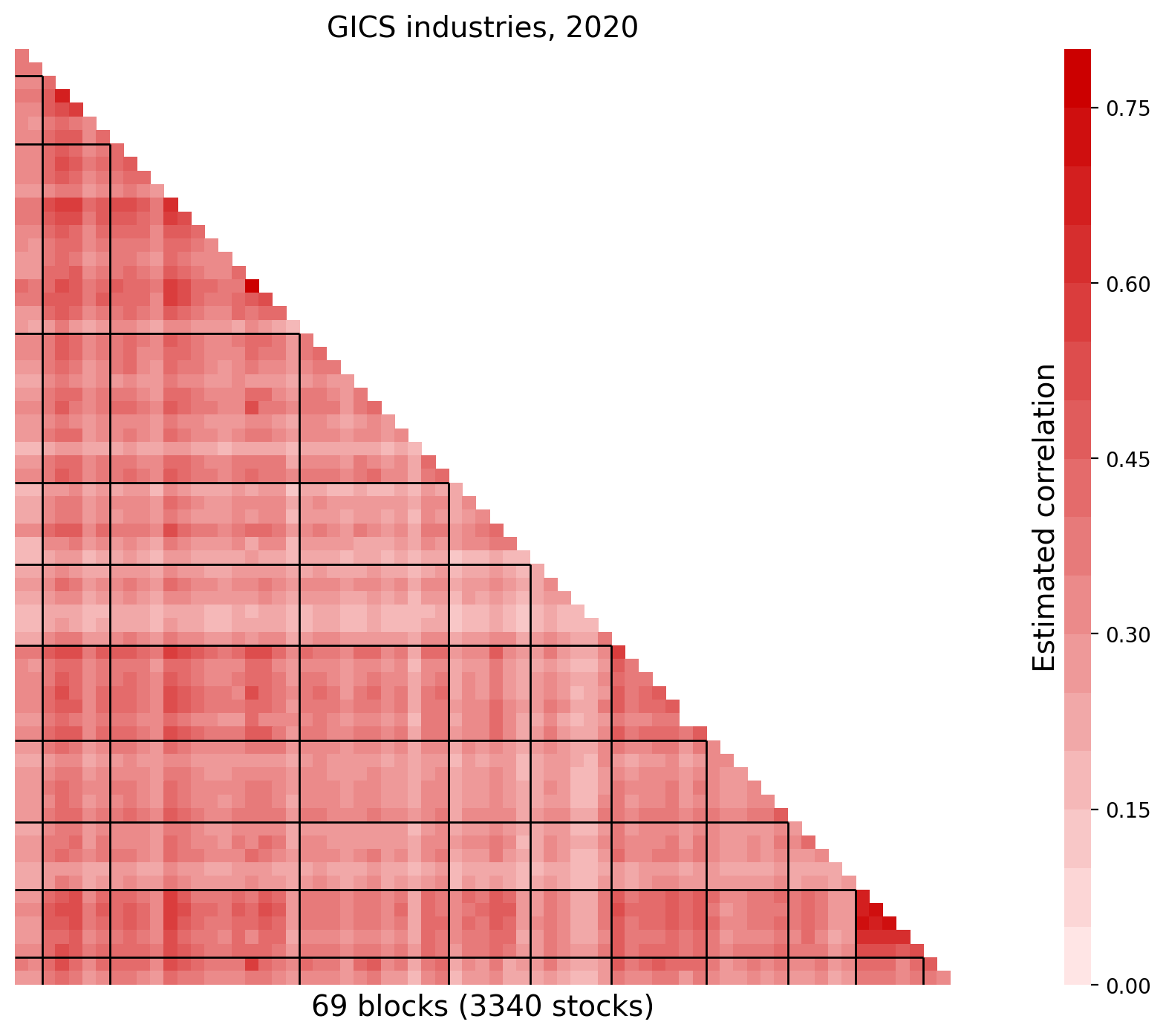}\tabularnewline
\end{tabular}}
\par\end{centering}
\begin{centering}
\subfloat[Sub-industry correlation structure ($K=152$ blocks)]{\centering{}%
\begin{tabular}{cc}
\includegraphics[scale=0.4]{Figures/crisp_mle_2019_sub} & \includegraphics[scale=0.4]{Figures/crisp_mle_2020_sub}\tabularnewline
\end{tabular}}
\par\end{centering}
\caption{(continued).}
\end{figure}

The estimated block correlation matrices based on Sectors, Groups,
Industries and Sub-industries are shown in Figure \ref{fig:BlockCorrelationsMatrices}.
Along the diagonal are the estimated correlation coefficients for
assets in the same block (within block correlations). Other estimates
are for pairs of assets from different blocks (between block correlations).
Due to the larger number of estimated correlation coefficients, we
present most estimates using color coding. A darker shade of red denotes
a stronger correlation.

From Figure \ref{fig:BlockCorrelationsMatrices} we can see that correlations
were generally higher in 2020 than in 2019. This can be attributed
to the COVID-19 Pandemic. When COVID-19 cases began to spread worldwide,
beyond isolated cases, the market experienced a large decline that
continued as lockdowns were imposed in most countries. The S\&P 500
index declined by more than 33\% from February 19, 2020 to March 23,
2020. This was followed by a strong rally where the market, from March
23, 2020 to the end of the year, increased by more than 63\%. The
block structure is somewhat more visible in 2020, which could be due
to the differentiated effect the pandemic had on different sectors
of the economy. For instance, in Figure \ref{fig:BlockCorrelationsMatrices}
we can see that the correlation between Utilities (55) and other sectors
increased substantially. Finance (40) tended to have the highest average
correlation with other sectors, but Utilities (55) had the highest
average correlation with other sectors in 2020. The low correlation
between Health Care (35) and other sectors is also very pronounced
in 2020. The blocks are listed in order of their GICS codes and we
include solid black lines to separate different sectors (the number
of blocks is too large to include labels for Industries and Sub-Industries
individually). The block partition based on Sub-Industries in panel
(d)\footnote{These are identical to the results presented in the Introduction in
Figure \ref{fig:Preview}.} reveals additional details about the correlation structure. Within
the Health Care sector (35) there is a distinct stripe that signifies
near zero correlations with all other blocks. Interestingly, this
stripe corresponds to two subindustries, Biotechnology (35201010)
and Pharmaceuticals (35202010). In the Figure \ref{fig:BlockCorrelationsMatrices}(d)
there is also a band with low correlations that is associated with
sub-industries in Materials (15), and these are Gold (15104030), Precious
Metals and Minerals (15104040), and Silver (15104045).
\begin{figure}[!ph]
\begin{centering}
\begin{tabular}{cc}
\includegraphics[scale=0.4]{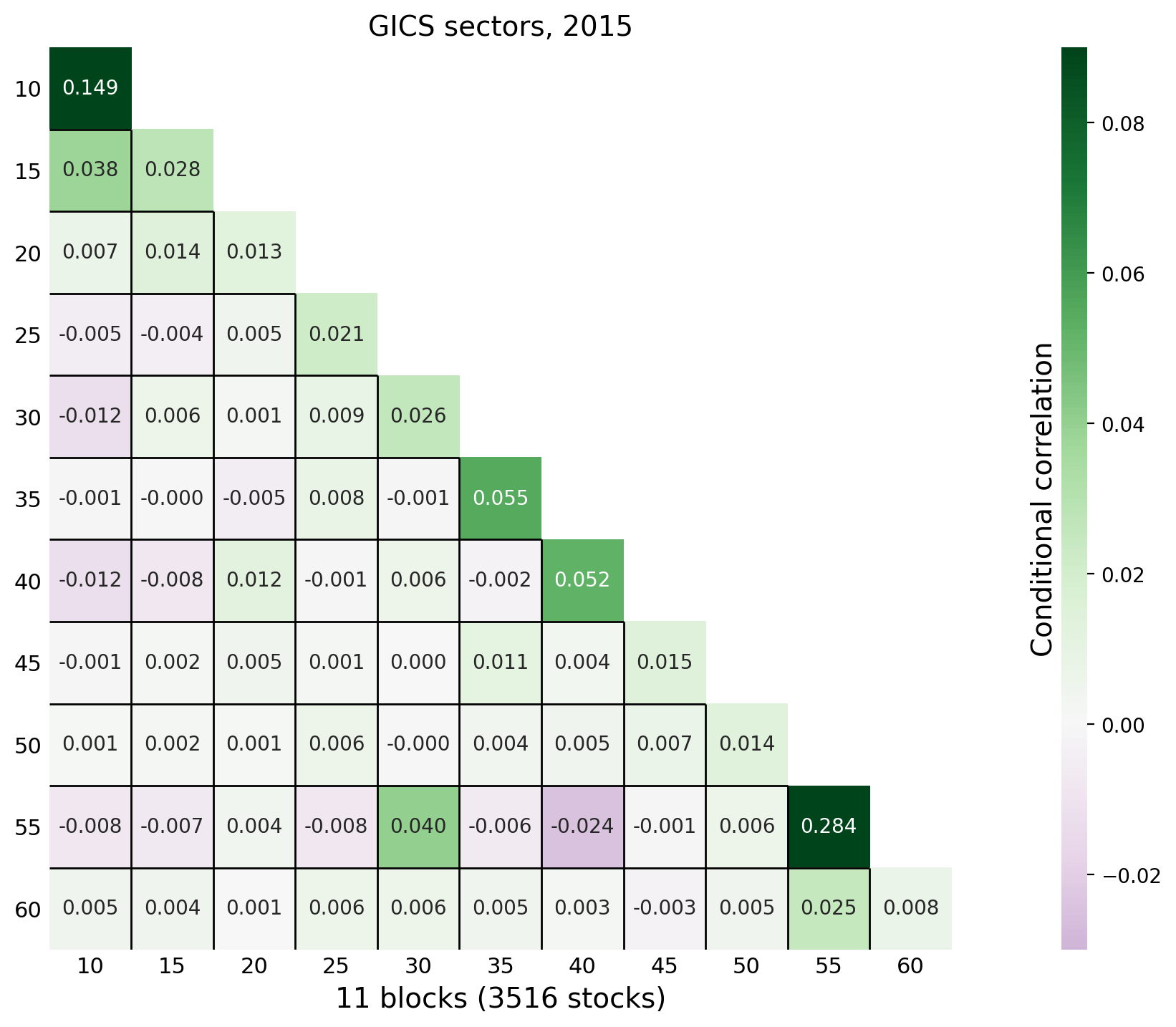} & \includegraphics[scale=0.4]{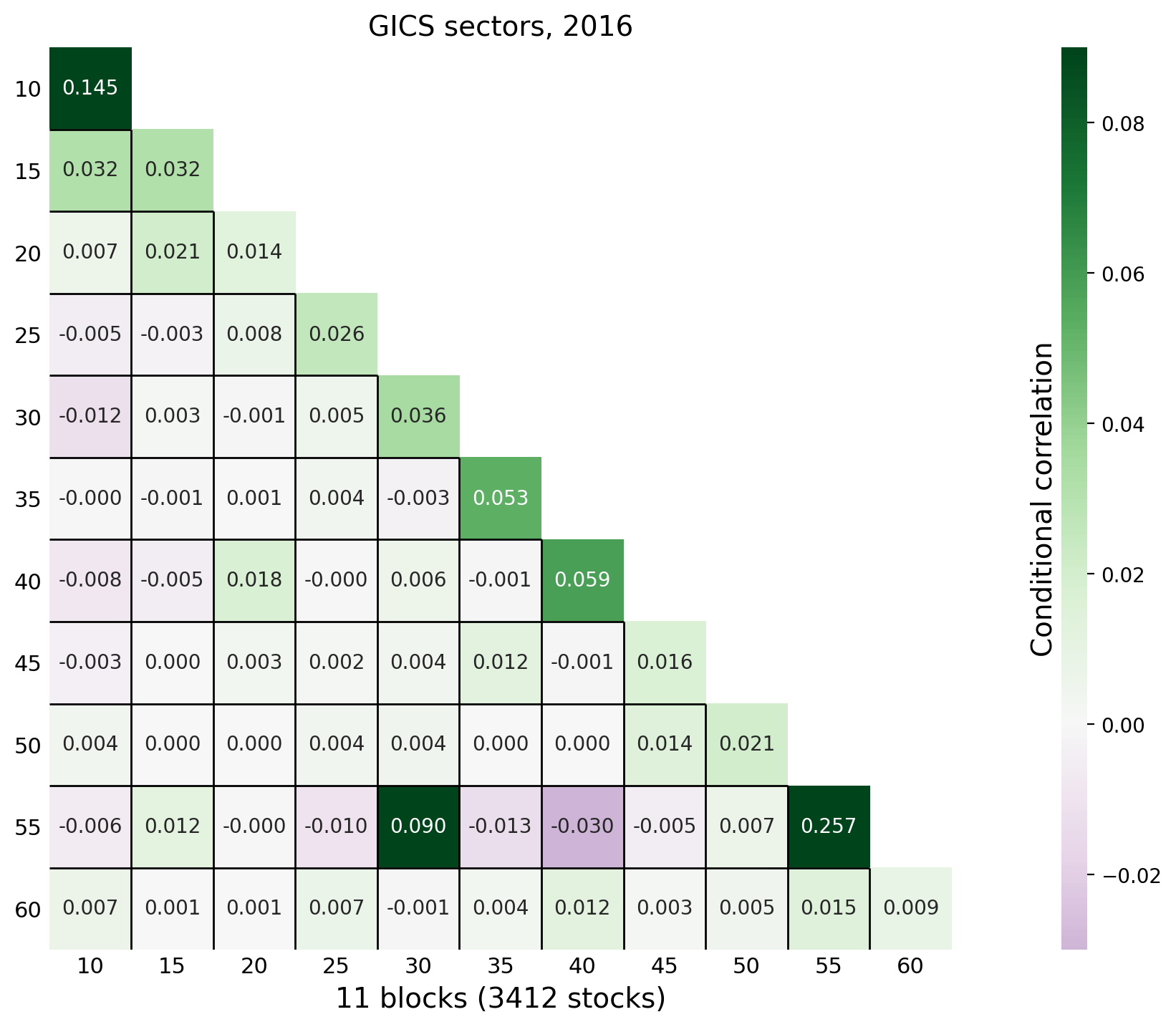}\tabularnewline
\includegraphics[scale=0.4]{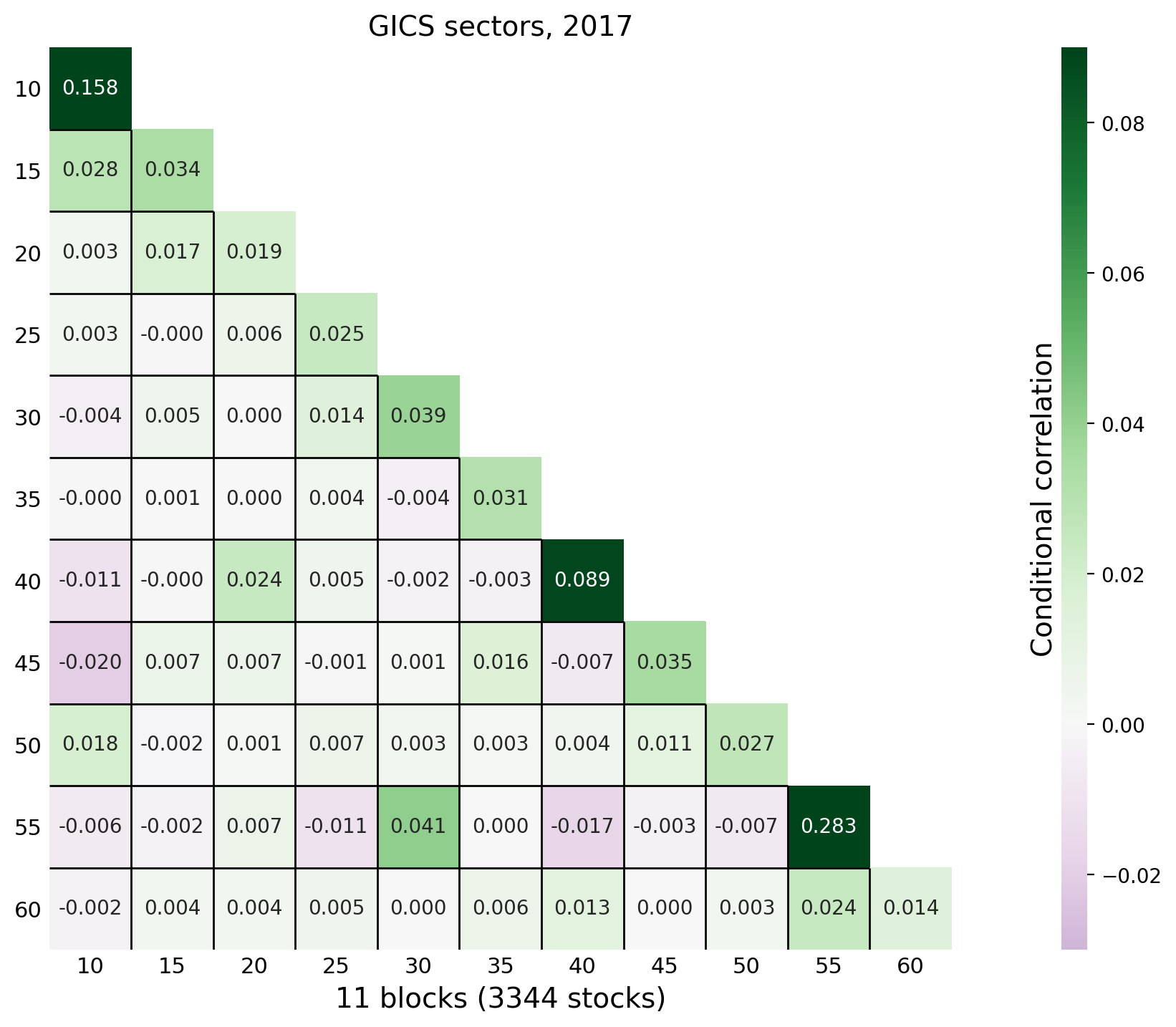} & \includegraphics[scale=0.4]{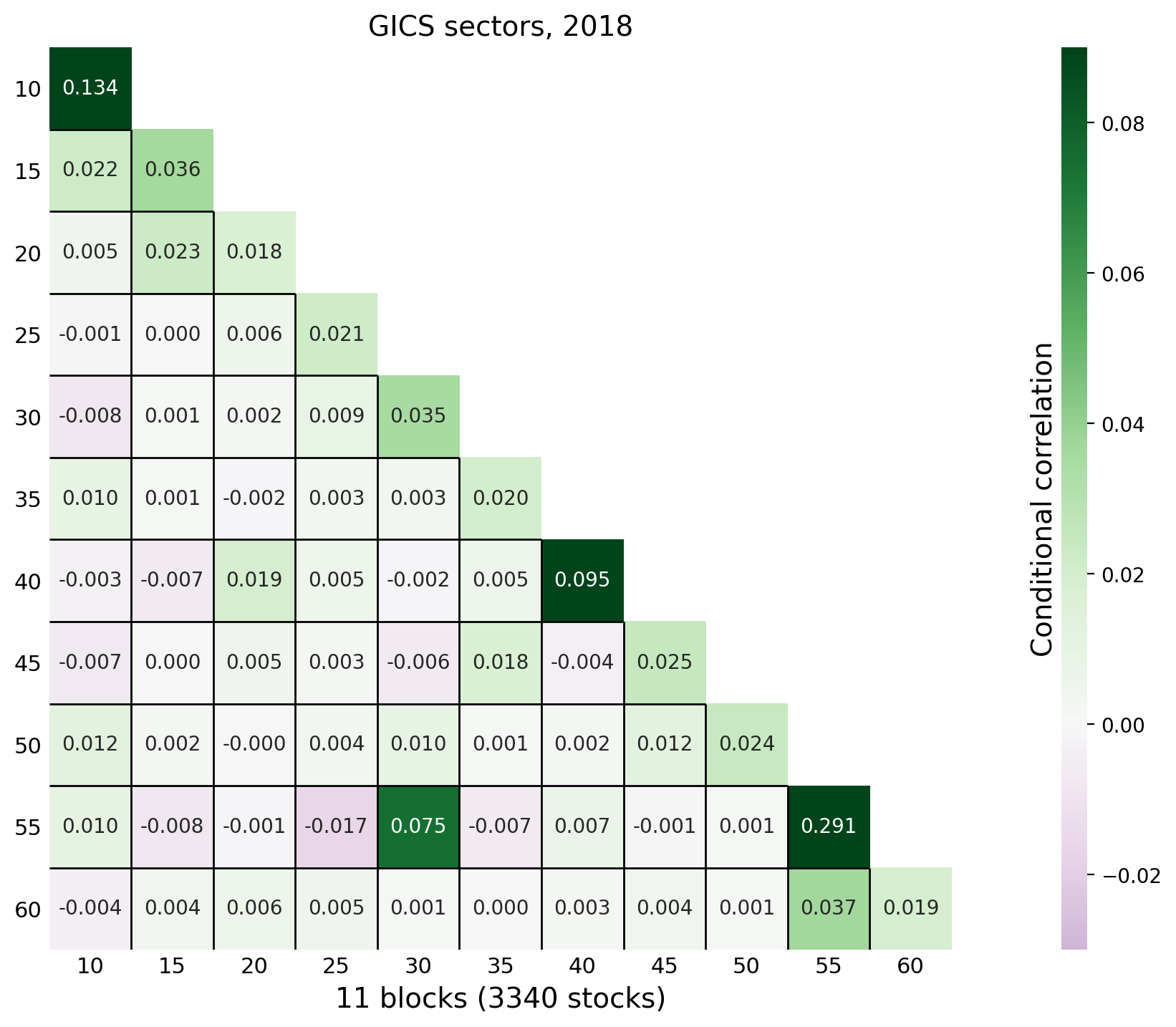}\tabularnewline
\includegraphics[scale=0.4]{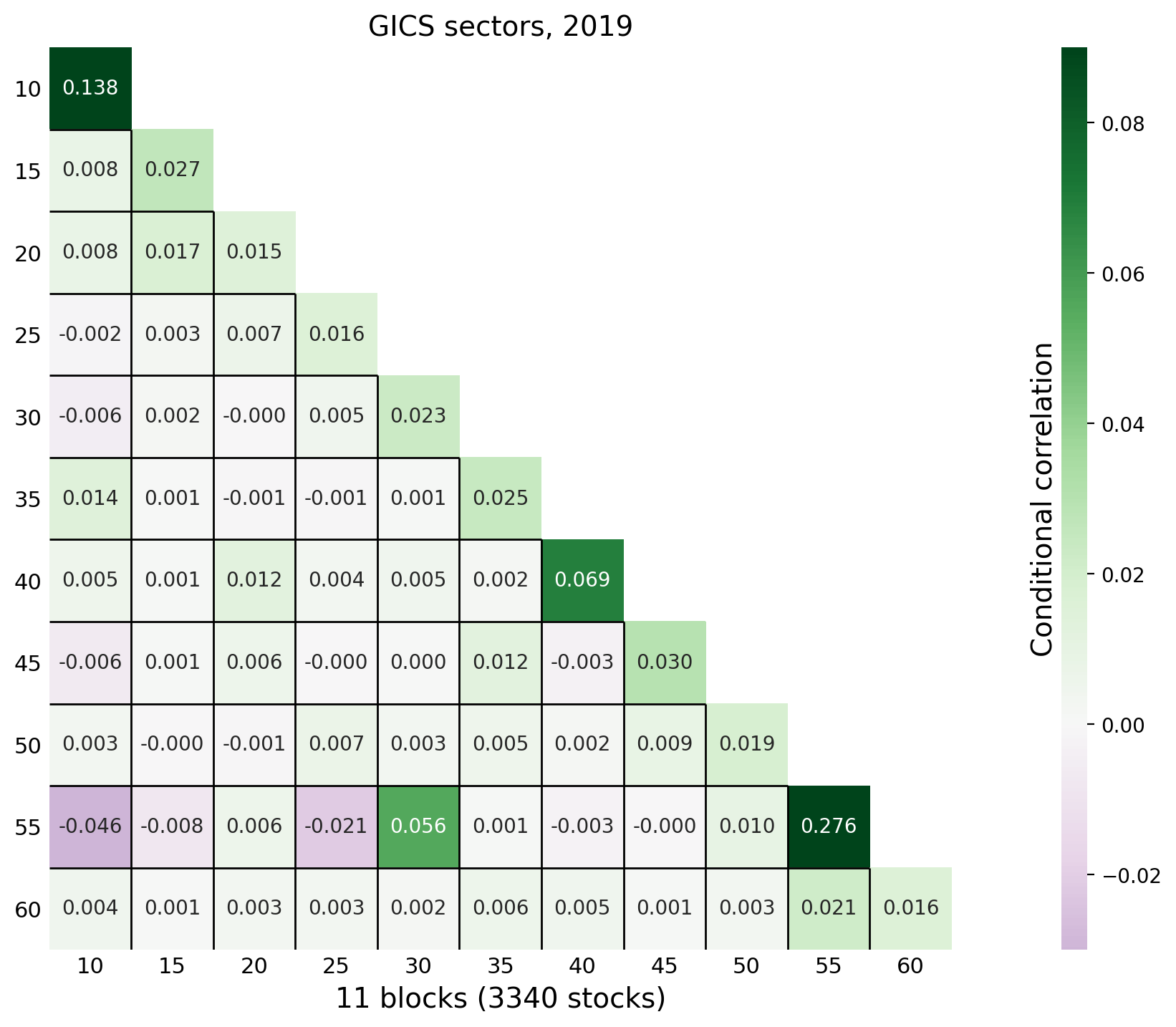} & \includegraphics[scale=0.4]{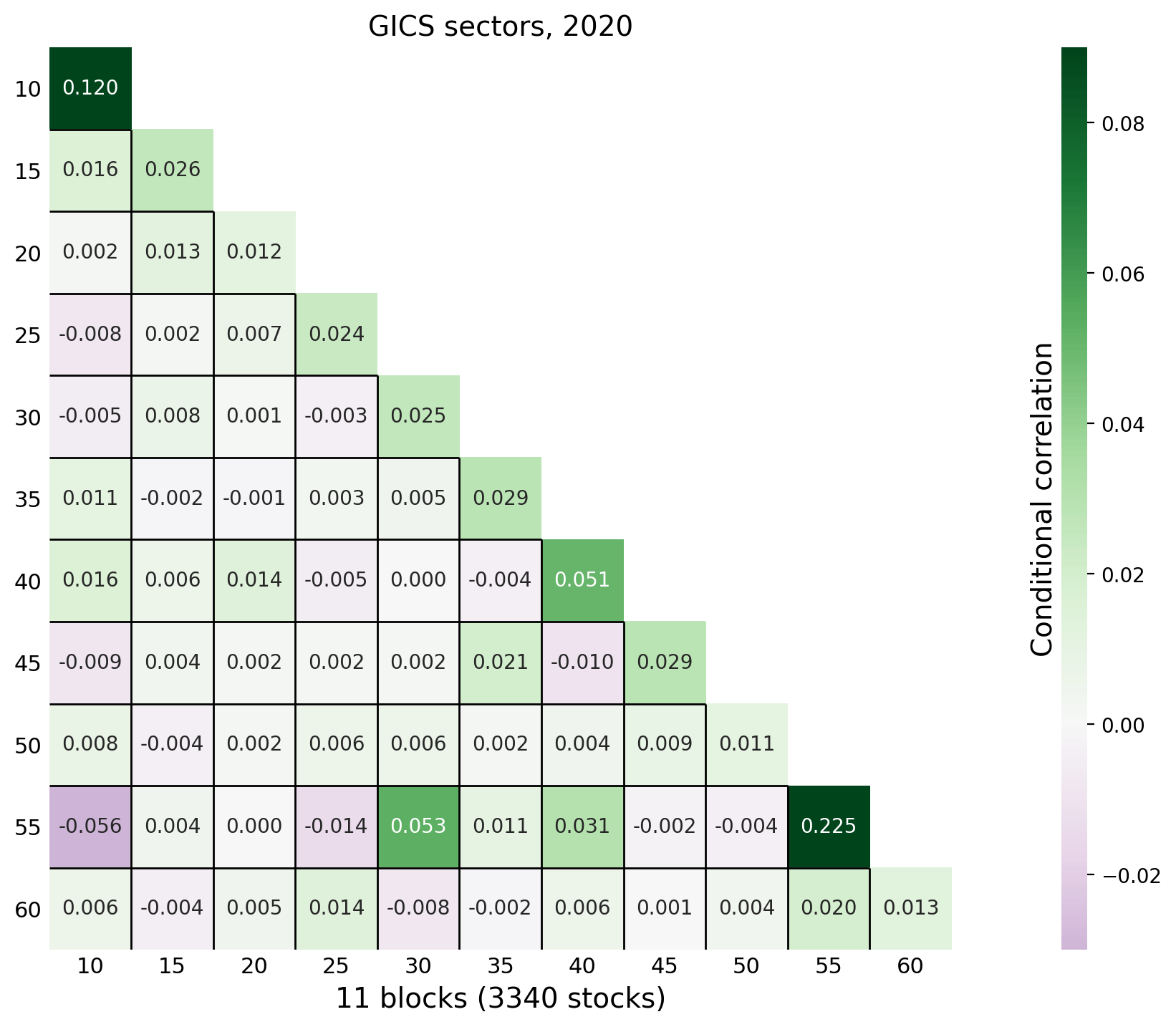}\tabularnewline
\end{tabular}
\par\end{centering}
\caption{Partial correlations for sector-block correlation matrix by calendar
year (2015 - 2020).\label{fig:PartialCorrelationsSector}}
\end{figure}

To further illustrate the usefulness of the block correlation structure,
we compute partial correlations for pairs of stocks. Partial correlations
require inversion of a high-dimensional matrix, a computation that
is greatly simplified with the block structure. In Figure \ref{fig:PartialCorrelationsSector}
we report the partial correlation for a pair of stocks, where we have
conditioned on all other stocks in other sectors. These partial correlations
are based on the estimated correlation matrices using the sector-block
structure.\footnote{The block structure greatly simplifies the computation of this type
of partial correlation. The formulae are derived and presented in
the Web Appendix.} Figure \ref{fig:PartialCorrelationsSector} includes results for
six calendar years, the results for all other calendar years are presented
in the Web Appendix.

One feature that stands out from the partial correlation matrices
is the similarity across calendar years, of which six years are shown
in Figure \ref{fig:PartialCorrelationsSector}. Had the annual estimates
of the correlation matrices been very noisy, we would not expect to
see very similar structures in estimates based on different data sets
(daily returns from different calendar years). The estimate from one
calendar year, tend to be similar to that of the neighboring years,
with some exceptions associated with the Global Financial Crisis and
the COVID-19 pandemic. This is precisely what we would expect if the
correlation structure is time-varying but typically evolves in a relatively
smooth manner. It is interesting that two sectors, Energy and Utilities,
have large degrees of residual correlation that are left unexplained
after having conditioned on all stocks in other sectors. This indicates
that these sectors need a sector specific factor to explain their
correlation structure. A potentially interesting application of the
partial correlation analysis, would be to extend the set of assets
with a set of ``factors'', such as the three Fama-French factors,
and other candidate factors. Computing partial correlations, where
the conditioning is on the factors, could be used to identify correlation
structures that are left unexplained by the factors. We leave this
for future research.
\begin{figure}[!ph]
\begin{centering}
\subfloat[Correlations and Partial Correlations for stock in Energy and Utilities
sectors]{\centering{}\includegraphics[scale=0.5]{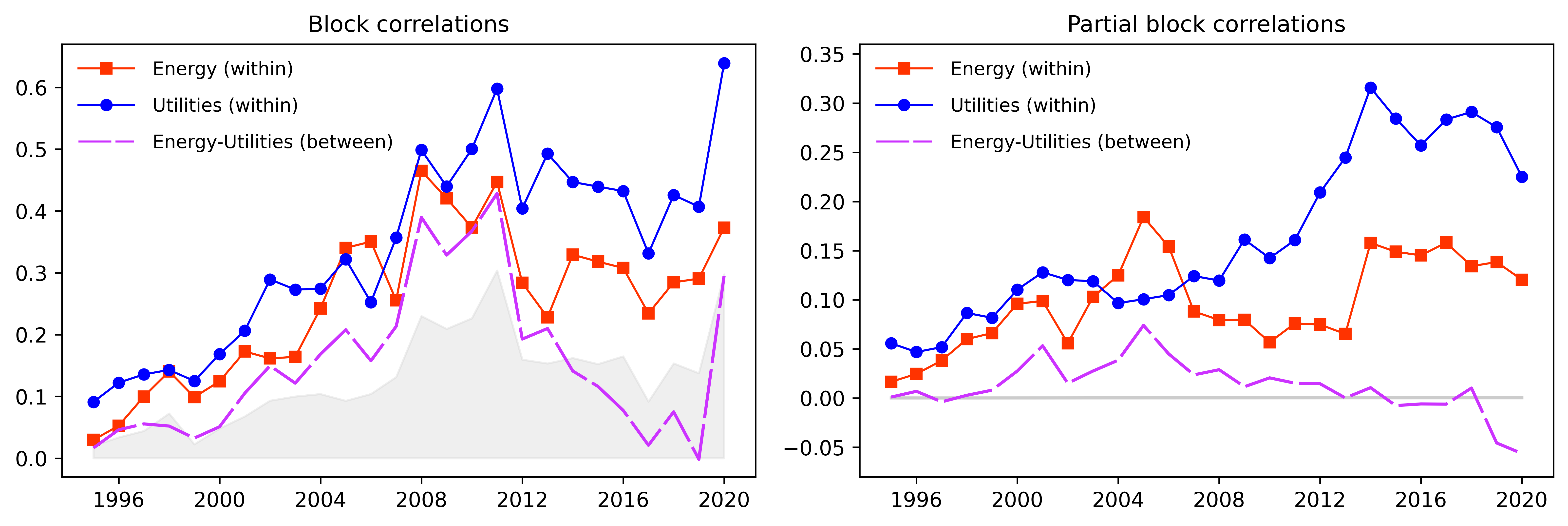}}
\par\end{centering}
\centering{}\subfloat[Bayesian Information Criterion (BIC)]{\centering{}\includegraphics[scale=0.54]{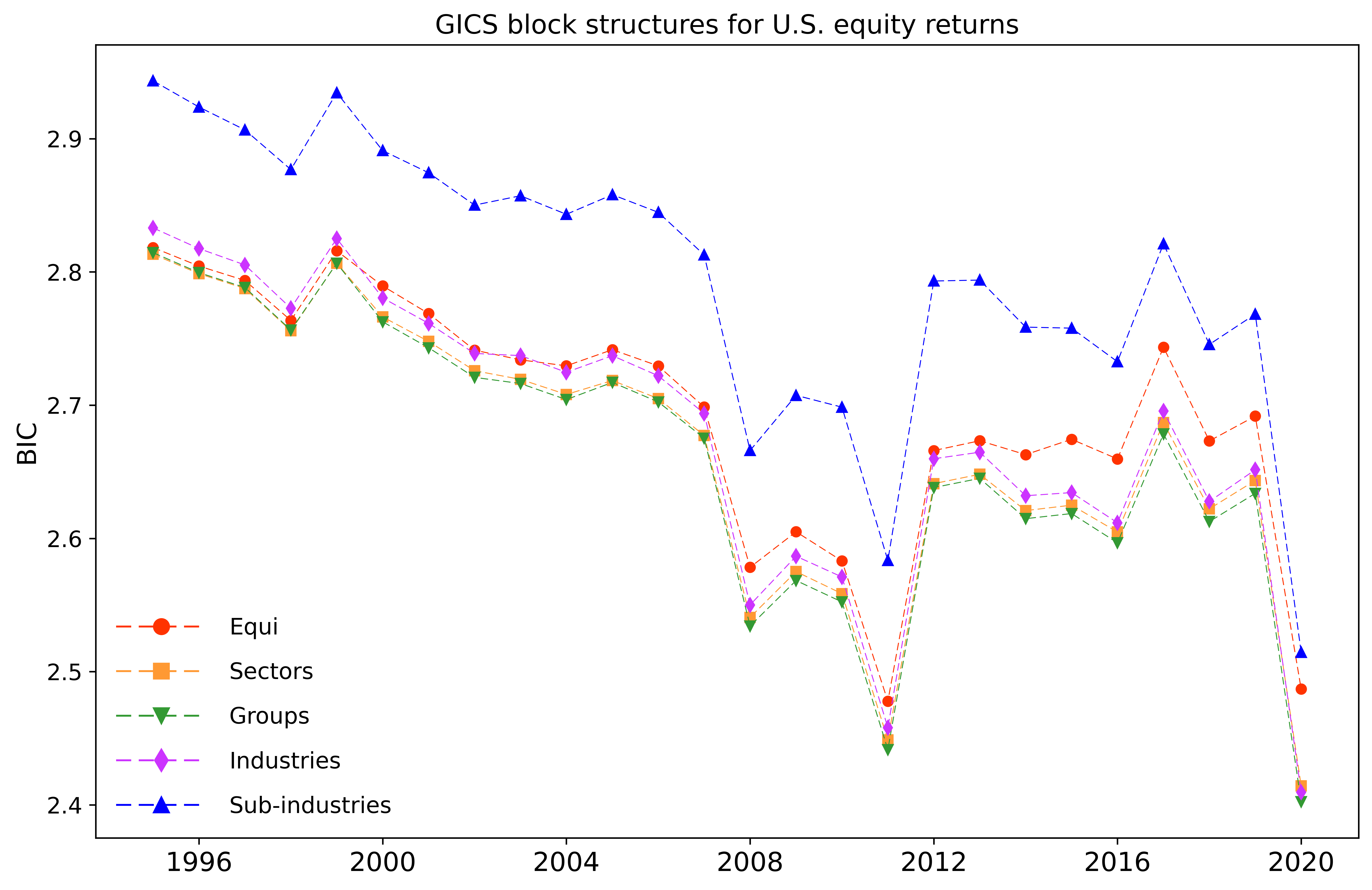}}\caption{Some selected empirical results for all calendar years.\label{fig:EnergyUtilityCorrs_BIC}}
\end{figure}

Figure \ref{fig:EnergyUtilityCorrs_BIC} presents selected results
for all 26 calendar years. In the upper panel (a), we present the
estimated correlations (left) and partial correlations (right) between
assets in the Energy and Utilities based on the sector-block structure.
The shaded areas represent the average correlation and average partial
correlation based on the equicorrelation structure for all assets.
Figure \ref{fig:EnergyUtilityCorrs_BIC} shows that the correlations
have been trending upwards and there has also been a great deal of
variation in the correlation between these two sectors. The partial
correlations are interesting, because they indicate that these two
sectors, Energy and Utilities, have large idiosyncratic components,
because a large fraction of the correlations between stocks within
either of these two sectors is unexplained by the thousands of stocks
in other sectors.

In the lower panel (b) of Figure \ref{fig:EnergyUtilityCorrs_BIC},
we present the BIC for each of the calendar years and each block structure.
Before 2000, the BIC always selected the block structures based on
sectors and after 2000 it systematically favors the block structure
based on Groups. The most heavily parametrized specification, which
is based on sub-industries, has the worst BIC in all calendar years.

\section{Concluding Remarks\label{sec:Concluding-Remarks}}

We have derived a canonical representation of block matrices. The
representation provides valuable simplifications for models with block
matrices, such as stochastic block models for large networks, and
models with block covariance and block correlation matrices. We derived
a number of expressions that greatly simplify the computation of the
Gaussian log-likelihood function with block covariance/correlation
matrices. We illustrated this in an empirical application, where we
estimate large covariance matrices for a vector with thousands of
assets, with daily returns over a single calendar year. Inverting
the covariance matrix, computing partial correlations, and evaluating
the Gaussian log-likelihood is straightforward once a block structure
is imposed.

The canonical representation and the related results are potentially
useful for regularizing large covariance matrices. For instance, one
could shrink the sample correlation matrix towards a block correlation
matrix, analogous to the way \citet{LedoitWolf:2004} proposed to
shrink towards the equicorrelation matrix with the MacGyver method,
see also \citet{Engle:2009}. This could possibly be extended to shrinkage
involving a convex combination of several block correlation matrices.

The canonical representation also paves new ways to testing block
structures in covariance and correlation matrices. This predominantly
amounts to testing a large number of zero-restrictions in the canonical
representation. We identified a number of transformations that preserves
the block structures, so testing of block structures could be based
on any of the transformations, rather than the original matrix. For
instance, block structures in a correlation matrix $C$ would be tested
on the canonical representation for $\log C$. This is potentially
interesting, because the connection between logarithmically transformed
correlation matrix and the Fisher transformation, see \citet{ArchakovHansen:Correlation}.
Finally, the group assignments, and hence $K$, will be unknown in
many empirical applications. The literature has therefore proposed
various classification methods to determine an appropriate block structure.
It is possible that the canonical representation will be useful for
this type of classification problems.

\appendix

\section*{Appendix of Proofs}

\noindent\textbf{Proof of Theorem \ref{thm:Canonical}.} For $k\neq l$,
we have $B_{[k,l]}=a_{kl}P_{[k,l]}$ if $a_{kl}=b_{kl}\sqrt{n_{k}n_{l}}$,
since the elements of $P_{[k,l]}$ are all equal to $\frac{1}{\sqrt{n_{k}n_{l}}}.$
For $k=l$, the diagonal elements differ from off-diagonal elements
by $\lambda_{k}=d_{k}-b_{kk}$, so that $B_{[k,k]}=b_{kk}n_{k}P_{[k,k]}+(d_{k}-b_{kk})I_{n_{k}}$.
Since $I_{n_{k}}=P_{[k,k]}+P_{[k,k]}^{\bot}$, we have $B_{[k,k]}=(b_{kk}n_{k}+d_{k}-b_{kk})P_{[k,k]}+(d_{k}-b_{kk})P_{[k,k]}^{\bot}=a_{kk}P_{[k,k]}+\lambda_{k}P_{[k,k]}^{\bot}$.
The canonical representation, (\ref{eq:Canonical}), follows by verifying
that $Q^{\prime}BQ$ is equal to the block-diagonal matrix in (\ref{eq:Canonical}).
This follows from the identities: $v_{n_{k}}^{\prime}P_{[k,l]}v_{n_{l}}=1$,
$v_{n_{k}}^{\prime}P_{[k,l]}v_{n_{l}\bot}=0$, $v_{n_{k}\bot}^{\prime}P_{[k,l]}v_{n_{l}\bot}=0$,
$v_{n_{k}}^{\prime}P_{[k,k]}^{\bot}v_{n_{k}}=0$, $v_{n_{k}}^{\prime}P_{[k,k]}^{\bot}v_{n_{k}\bot}=0$,
and $v_{n_{k}\bot}^{\prime}P_{[k,k]}^{\bot}v_{n_{k}\bot}=I_{n_{k}-1}$,
and the fact that $Q^{\prime}Q=I_{n}$, so that $Q^{-1}=Q^{\prime}$,
and hence $B=QQ^{\prime}BQQ^{\prime}$. This proves (\ref{eq:Canonical}).
$\square$

\noindent\textbf{Proof of Corollary \ref{cor:BlockCovMatrix}.} The
first result for the eigenvalues of $B$ and the determinant of $B$,
follows immediately from (\ref{eq:Canonical}). The results for $f(B)$,
where $f$ denotes the $q$-th power of a matrix, the matrix exponential,
or the matrix logarithm, follow by $f(B)=Qf(D)Q^{\prime}$ and using
the structure in $Q$, such as $v_{n_{k}}v_{n_{l}}^{\prime}=P_{[k,l]}$
and $v_{n_{k}\bot}v_{n_{k}\bot}^{\prime}=P_{[k,k]}^{\bot}$. This
completes the proof. $\square$

\noindent\textbf{Proof of Theorem \ref{thm:RetangularBlockConsistent}.}
Since $V_{0,t}W_{0,t}^{\prime}$ and $V_{k,t}^{\prime}W_{k,t}/(n_{k}-1)$
are stationary and ergodic with expected values $A$ and $\lambda_{k}$,
it follows from the law of large number for ergodic processes. Thus,
$\hat{A}$ is consistent for $A$ and $\hat{\lambda}_{k}$ is consistent
for $\lambda_{k}$, $k=1,\ldots,K$, and hence, $Q\hat{D}_{z\tilde{x}}Q^{\prime}\overset{p}{\rightarrow}QD_{z\tilde{x}}Q^{\prime}=\Sigma_{z\tilde{x}}$.
$\square$

\noindent\textbf{Proof of Corollary \ref{cor:BlockPreservingTrans}.}
It follows from Theorem \ref{thm:Canonical} and Corollary \ref{cor:BlockCovMatrix}
by setting $d_{k}=1$ for all $k$. Some expressions can also be verified
directly. For instance, one can verify the expression for $C^{-1}$
, by noting that diagonal blocks of $C^{-1}$ are given by 
\[
(C^{-1})_{[k,k]}=\sum_{m=1}^{K}a_{km}P_{[k,m]}a_{mk}^{\#}P_{[m,k]}+(1-\rho_{kk})P_{[k,k]}^{\bot}\tfrac{1}{1-\rho_{kk}}P_{[k,k]}^{\bot}=\sum_{m=1}^{K}a_{km}a_{mk}^{\#}P_{[k,k]}+P_{[k,k]}^{\bot}=I,
\]
where we used that $a_{mk}^{\#}$ are the elements of the $A^{-1}$
so we have $\sum_{m=1}^{K}a_{km}a_{mk}^{\#}=1$. Next, for $k\neq m$,
we have
\[
(C^{-1})_{[k,l]}=\sum_{m=1}^{K}a_{km}P_{[k,m]}a_{mk}^{\#}P_{[m,l]}+\frac{a_{kl}}{b_{l}}P_{[k,l]}P_{[l,l]}^{\bot}+b_{k}a_{kl}^{\#}P_{[k,k]}^{\bot}P_{[k,l]}=\sum_{m=1}^{K}a_{km}a_{ml}^{\#}P_{[k,l]}=0,
\]
where we used that $P_{[k,m]}P_{[m,l]}=P_{[k,l]}$ and $P_{kl}P_{[l,l]}^{\bot}=P_{[k,l]}(I_{s_{l}}-P_{[l,l]})=0$,
and that $\sum_{m=1}^{K}a_{km}a_{ml}^{\#}=0$, for $k\neq l$. This
completes the proof. $\square$

\noindent\textbf{Proof of Theorem \ref{thm:BlockVarianceMLE}.} The
expression, (\ref{eq:-2logL}), shows that the log-likelihood function
is made up of two terms:
\[
-2N\left[\log\det A+\mathrm{tr}\{A^{-1}\tfrac{1}{T}\sum_{t=1}^{T}Y_{0,t}Y_{0,t}^{\prime}\}\right],
\]
and 
\[
-2N\sum_{k=1}^{K}(n_{k}-1)\left(\log\lambda_{k}+\frac{\tfrac{1}{T}\sum_{t=1}^{T}\frac{Y_{k,t}^{\prime}Y_{k,t}}{n_{k}-1}}{\lambda_{k}}\right).
\]
It is well known that $X=\arg\min_{\Theta}\log\det\Theta+\mathrm{tr}\{\Theta^{-1}X\}$,
such that $\hat{A}=\tfrac{1}{T}\sum_{t=1}^{T}Y_{0,t}Y_{0,t}^{\prime}$
maximizes the first term and that $\hat{\lambda}_{k}=\tfrac{1}{T}\sum_{t=1}^{T}\frac{Y_{k,t}^{\prime}Y_{k,t}}{n_{k}-1}$
maximizes the elements of the second term. Since $(A,\lambda_{1},\ldots,\lambda_{K})$
is merely a reparameterization of the elements of the block covariance
matrix $\Sigma$, it follows that $\hat{\Sigma}=Q\hat{D}Q^{\prime}$
is the maximum likelihood estimator of $\Sigma$. It is easy to verify
that this result is also valid in the special case, where one or more
of the blocks are 1-dimensional. In this case, $\sigma_{kk}$ is undefined,
and so is $\hat{\lambda}_{k}$, while $\sigma_{k}^{2}$ is identified
from the corresponding diagonal element of $A$, since $\hat{a}_{kk}=\sigma_{k}^{2}$,
when $n_{k}=1$. $\square$

\noindent\textbf{Proof of Corollary \ref{cor:BlockCorrelationMLE}.}
Define the sample covariance matrix, $S=\{s_{ij}\}_{i,j=1}^{n}=\frac{1}{T}\sum_{t=1}^{T}X_{t}X_{t}^{\prime}$,
and $\eta_{i}=\sqrt{s_{i}^{2}/\sigma_{i}^{2}}$, where $s_{i}^{2}\equiv s_{ii}$
is the sample variance for $X_{i,t}$, $i=1,\ldots,n$. Next define
the the matrices $R$ and $M$ with elements
\[
r_{ij}=\frac{s_{ij}}{\sqrt{s_{ii}s_{jj}}},\qquad m_{ij}=\frac{s_{ij}}{\sigma_{i}\sigma_{j}}=r_{ij}\eta_{i}\eta_{j},\qquad i,j=1,\ldots,n,
\]
respectively. Let $R_{[k,l]}$ and $M_{[k,l]}$ be the $n_{k}\times n_{l}$
submatrix of $R$ and $M$, respectively, that corresponds to the
$(k,l)$-th block, and let $\eta_{[k]}=\{\eta_{i\in\mathcal{I}_{k}}\}$
be the subvector of $\eta$, with the elements associated with the
$k$-th block.

The log-likelihood function is proportional to,
\[
-\frac{2}{T}\ell(\Sigma)=\log|\Sigma|+\tfrac{1}{T}\sum_{t=1}^{T}X_{t}^{\prime}\Sigma^{-1}X_{t}=\sum_{j=1}^{n}\log\sigma_{j}^{2}+\log|C|+\mathrm{tr}\{C^{-1}M\},
\]
where
\begin{eqnarray*}
\mathrm{tr}\{C^{-1}M\} & = & \mathrm{tr}\{Q^{\prime}C^{-1}QQ^{\prime}MQ\},\\
 & = & \mathrm{tr}\{A^{-1}V(\eta)\}+\sum_{k=1}^{K}\frac{1}{\lambda_{k}}\mathrm{tr}\{M_{[k,k]}(I-v_{n_{k}}v_{n_{k}}^{\prime})\}\\
 & = & \mathrm{tr}\{A^{-1}V(\eta)\}+\sum_{k=1}^{K}\frac{1}{\lambda_{k}}(\eta_{[k]}^{\prime}\eta_{[k]}-[V(\eta)]_{k,k}),
\end{eqnarray*}
with $[V(\eta)]_{k,l}=\tfrac{1}{\sqrt{n_{k}n_{l}}}\eta_{[k]}^{\prime}R_{[k,l]}\eta_{[l]}$.
Thus, the log-likelihood can be expressed as
\begin{eqnarray*}
-\tfrac{T}{2}\ell(\eta,A,\lambda) & \propto & -\sum_{i=1}^{n}\log\eta_{i}^{2}+\log|A|+\mathrm{tr}\{A^{-1}V(\eta)\},\\
 &  & +\sum_{k=1}^{K}(n_{k}-1)\log\lambda_{k}+\frac{\eta_{[k]}^{\prime}\eta_{[k]}-[V(\eta)]_{k,k}}{\lambda_{k}}.
\end{eqnarray*}
Next 
\begin{eqnarray*}
\tilde{A}(\eta) & = & V(\eta)=\arg\max_{A}\left(\log|A|+\mathrm{tr}\{A^{-1}V(\eta)\}\right),\\
\tilde{\lambda}_{k}(\eta) & = & \frac{\eta_{[k]}^{\prime}\eta_{[k]}-[V(\eta)]_{k,k}}{n_{k}-1}=\arg\max\left((n_{k}-1)\log\lambda_{k}+\frac{\eta_{[k]}^{\prime}\eta_{[k]}-[V(\eta)]_{k,k}}{\lambda_{k}}\right),
\end{eqnarray*}
but $\tilde{A}(\eta)$ and $\tilde{\lambda}_{k}(\eta)$, $k=1,\ldots,K$,
need not satisfy their cross restrictions, which by Theorem \ref{thm:Canonical}
(set $d_{k}=1$) are given by 
\[
\lambda_{k}=\frac{n_{k}-a_{kk}}{n_{k}-1},\qquad k=1,\ldots,K.
\]
However, it follows that the log-likelihood is bounded by $-\tfrac{T}{2}\ell(\eta,A,\lambda)\leq-\tfrac{T}{2}\ell(\eta,\tilde{A}(\eta),\tilde{\lambda}(\eta))$,
and if we define $\delta_{k}=\sqrt{\eta_{[k]}^{\prime}\eta_{[k]}/n_{k}}$
and $\tilde{\eta}_{[k]}=\delta_{k}^{-1}\eta_{[k]}$, which has $\mathbb{\tilde{\eta}}_{[k]}^{\prime}\mathbb{\tilde{\eta}}_{[k]}=n_{k}$,
then $-\tfrac{T}{2}\ell(\eta,\tilde{A}(\eta),\tilde{\lambda}(\eta))$
equals
\begin{eqnarray*}
-\tfrac{T}{2}\ell(\mathbb{\tilde{\eta}},\delta,\tilde{A}(\eta),\tilde{\lambda}(\eta)) & \propto & -\sum_{k}n_{k}\log\delta_{k}^{2}-\sum_{i=1}^{n}\log\tilde{\eta}_{i}^{2}+\sum_{k}\log\delta_{k}^{2}+\log|V(\tilde{\eta})|\\
 &  & +\sum_{k}(n_{k}-1)\log\left(\delta_{k}^{2}\frac{\tilde{\eta}_{[k]}^{\prime}\tilde{\eta}_{[k]}-[V(\tilde{\eta})]_{k,k}}{n_{k}-1}\right)\\
 & = & -\sum_{i=1}^{n}\log\tilde{\eta}_{i}^{2}+\log|V(\tilde{\eta})|+\sum_{k}(n_{k}-1)\log\frac{\tilde{\eta}_{[k]}^{\prime}\tilde{\eta}_{[k]}-[V(\tilde{\eta})]_{k,k}}{n_{k}-1}.
\end{eqnarray*}
Conveniently, this expression does not depend on $\delta$, such that
$\ell(\mathbb{\tilde{\eta}},\delta,\tilde{A}(\eta),\tilde{\lambda}(\eta))=\ell(\mathbb{\tilde{\eta}},\iota_{K},\tilde{A}(\tilde{\eta}),\tilde{\lambda}(\tilde{\eta}))$,
where $\iota_{K}=(1,\ldots,1)^{\prime}\in\mathbb{R}^{K}$ and if we
set $\delta_{k}=1$ for all $k$, then the cross restrictions are
$\tilde{\lambda}_{k}(\eta)=(n_{k}-[V(\eta)]_{k,k})/(n_{k}-1)$ and
the cross restrictions are satisfied. So, without loss of generality
we can assume that $\mathbb{\eta}_{[k]}^{\prime}\mathbb{\eta}_{[k]}=n_{k}$
for all $k$ ($\delta_{k}=1$) and it follows that 
\[
\hat{\rho}_{k}=1-\hat{\lambda}_{k}=1-\frac{n_{k}-[V(\eta)]_{k,k}}{n_{k}-1}=\frac{[V(\eta)]_{k,k}-1}{n_{k}-1}=\frac{\tfrac{1}{n_{k}}\eta_{[k]}^{\prime}R_{[k,k]}\eta_{[k]}-1}{n_{k}-1}=\frac{\eta_{[k]}^{\prime}(R_{[k,k]}-I_{k})\eta_{[k]}}{n_{k}(n_{k}-1)},
\]
where we used $\eta_{[k]}^{\prime}\eta_{[k]}=n_{k}$ in the last identity.
This shows that, $\hat{\rho}_{k}$ is a weighted average of the empirical
correlations in the $k$-th diagonal block, with equal weighting in
the special case where $\eta_{[k]}=\iota_{n_{k}}$.

The remaining optimization problem is to maximizing the concentrated
log-likelihood which amounts to $\min_{\eta}f(\eta)$ subject to $\eta_{[k]}^{\prime}\eta_{[k]}=n_{k}$,
for $k=1,\ldots,K$, where
\[
f(\eta)=-\sum_{i=1}^{n}\log\eta_{i}^{2}+\log|V(\eta)|+\sum_{k}(n_{k}-1)\log\left(1+\frac{\eta_{[k]}^{\prime}(R_{[k,k]}-I_{k})\eta_{[k]}}{n_{k}(n_{k}-1)}\right).
\]
Let $\hat{\eta}$ denote the solution to this problem, then $\hat{\sigma}_{i}^{2}=\hat{\eta}_{i}^{2}s_{i}^{2}$
is the maximum likelihood estimator of $\sigma_{i}^{2}$, $i=1,\ldots,n$.
$\square$

Notice that 
\begin{eqnarray*}
f(\eta) & \simeq & -\sum_{i=1}^{n}\log\eta_{i}^{2}+\log|V(\eta)|+\sum_{k}(n_{k}-1)\frac{\eta_{[k]}^{\prime}(R_{[k,k]}-I_{k})\eta_{[k]}}{n_{k}(n_{k}-1)}.\\
 & = & -\sum_{i=1}^{n}\log\eta_{i}^{2}+\log|V(\eta)|+\sum_{k}\tfrac{\eta_{[k]}^{\prime}(R_{[k,k]}-I_{k})\eta_{[k]}}{n_{k}}\\
 & = & -\sum_{i=1}^{n}\log\eta_{i}^{2}+\log|V(\eta)|+\mathrm{tr}\{V(\eta)\}-K.
\end{eqnarray*}

\noindent\textbf{Proof of Proposition \ref{prop:Score}.} Recall
that $a_{kk}=\sigma_{k}^{2}+(n_{k}-1)\sigma_{kk}$, $a_{kl}=\sigma_{kl}\sqrt{n_{k}n_{l}}$,
for $k\neq l$, and $\lambda_{k}=\sigma_{k}^{2}-\sigma_{kk}$. It
follows that 
\[
\frac{\partial(\log\det A+y_{0}^{\prime}A^{-1}y_{0})}{\partial a_{kl}}=\mathrm{tr}\{A^{-1}(e_{k}e_{l}^{\prime})(I-A^{-1}y_{0}y_{0}^{\prime})=e_{l}^{\prime}(I-A^{-1}y_{0}y_{0}^{\prime})A^{-1}e_{k}=M_{l,k},
\]
where $M=A^{-1}-A^{-1}y_{0}y_{0}^{\prime}A^{-1}$. From the expression
(\ref{eq:-2logL}), we find
\begin{eqnarray*}
\frac{\partial(-2\ell)}{\partial\sigma_{k}^{2}} & = & \frac{\partial(\log\det A+y_{0}^{\prime}A^{-1}y_{0})}{\partial a_{kk}}+(\tfrac{n_{k}-1}{\lambda_{k}}-\tfrac{y_{k}^{\prime}y_{k}}{\lambda_{k}^{2}})=M_{k,k}+(\tfrac{n_{k}-1}{\lambda_{k}}-\tfrac{y_{k}^{\prime}y_{k}}{\lambda_{k}^{2}})\\
\frac{\partial(-2\ell)}{\partial\sigma_{kk}} & = & (n_{k}-1)\frac{\partial(\log\det A+y_{0}^{\prime}A^{-1}y_{0})}{\partial a_{kk}}-\left(\tfrac{n_{k}-1}{\lambda_{k}}-\tfrac{y_{k}^{\prime}y_{k}}{\lambda_{k}^{2}}\right)=n_{k}M_{k,k}-\frac{\partial(-2\ell)}{\partial\sigma_{k}^{2}},
\end{eqnarray*}
and, for $k\neq l$, we find that
\[
\frac{\partial(-2\ell)}{\partial\sigma_{kl}}=\sqrt{n_{k}n_{l}}\left(\frac{\partial(\log\det A+y_{0}^{\prime}A^{-1}y_{0})}{\partial a_{kl}}+\frac{\partial(\log\det A+y_{0}^{\prime}A^{-1}y_{0})}{\partial a_{lk}}\right)=2\sqrt{n_{k}n_{l}}M_{k,l},
\]
where we used that $M$ is symmetric. $\square$

{\footnotesize{}\clearpage\bibliographystyle{agsm}
\bibliography{prh}
}{\footnotesize\par}
\end{document}

%% file: crisp_mle_stat_4.tex
\scriptsize
\begin{tabularx}{\textwidth}{ >{\hsize=2.1\hsize}X  Y  Y  >{\hsize=0.96\hsize}X  >{\hsize=0.96\hsize}X  >{\hsize=0.96\hsize}X  >{\hsize=0.96\hsize}X  >{\hsize=0.96\hsize}X  >{\hsize=1\hsize}X  Y  >{\hsize=0.3\hsize}X  >{\hsize=0.8\hsize}X  }
\\[-1.0cm]
\toprule
\midrule
  &  \multicolumn{7}{c}{Summary statistics of estimated block correlations} &  &  & \multicolumn{2}{c}{\# blocks}  \\
\cmidrule(l){2-3}
\cmidrule(l){4-8}
\cmidrule(l){11-12}
 Block & \multicolumn{1}{c}{Mean}  & \multicolumn{1}{c}{Std.}  & \multicolumn{1}{c}{Min}  & \multicolumn{1}{c}{$Q_{10\%}$}  & \multicolumn{1}{c}{$Q_{50\%}$}  & \multicolumn{1}{c}{$Q_{90\%}$}  & \multicolumn{1}{c}{Max}  & \multicolumn{1}{c}{$-\frac{2\ell}{nT}$}  & \multicolumn{1}{c}{$\frac{1}{nT}$BIC}  & \multicolumn{1}{c}{$K$}  & \multicolumn{1}{c}{$\frac{K(K+1)}{2}$}  \\
 structure & \multicolumn{11}{c}{}  \\
\midrule
\\[-0.1cm]
\multicolumn{1}{l}{\textit{}} & \multicolumn{5}{l}{\textit{U.S. market in 2019 (3340 stocks and 252 days)}} &  &  &  &  &  & \\
\\[-0.1cm]
 Equicorrelation & 0.138 & 0  & 0.138 & 0.138 & 0.138 & 0.138 & 0.138 & 2.69177 & 2.69178 & 1 & 1 \\
 Sectors & 0.134 & 0.062 & -0.002 & 0.072 & 0.127 & 0.208 & 0.407 & 2.64243 & 2.64349 & 11 & 66 \\
 Groups & 0.135 & 0.052 & -0.002 & 0.080 & 0.129 & 0.201 & 0.407 & 2.62892 & 2.63378 & 24 & 300 \\
 Industries & 0.137 & 0.063 & -0.059 & 0.074 & 0.129 & 0.212 & 0.594 & 2.61240 & 2.65152 & 69 & 2415 \\
 Sub-industries & 0.143 & 0.082 & -0.147 & 0.058 & 0.133 & 0.247 & 0.789 & 2.57997 & 2.76832 & 152 & 11628 \\
\\[0.0cm]
\multicolumn{1}{l}{\textit{}} & \multicolumn{5}{l}{\textit{U.S. market in 2020 (3340 stocks and 253 days)}} &  &  &  &  &  & \\
\\[-0.1cm]
 Equicorrelation & 0.298 & 0  & 0.298 & 0.298 & 0.298 & 0.298 & 0.298 & 2.48686 & 2.48687 & 1 & 1 \\
 Sectors & 0.317 & 0.079 & 0.191 & 0.223 & 0.318 & 0.415 & 0.639 & 2.41314 & 2.41421 & 11 & 66 \\
 Groups & 0.309 & 0.071 & 0.153 & 0.212 & 0.312 & 0.395 & 0.639 & 2.39740 & 2.40224 & 24 & 300 \\
 Industries & 0.327 & 0.083 & 0.134 & 0.220 & 0.321 & 0.433 & 0.763 & 2.37039 & 2.40938 & 69 & 2415 \\
 Sub-industries & 0.329 & 0.112 & -0.018 & 0.191 & 0.324 & 0.472 & 0.913 & 2.32680 & 2.51444 & 152 & 11628 \\
\\[-0.2cm]
\midrule
\bottomrule
\end{tabularx}